\documentclass[usenatbib]{mn2e}
\usepackage[pdftex]{graphicx}
\newcommand{\be}{\begin{equation}}
\newcommand{\ee}{\end{equation}}
\newcommand{\bea}{\begin{eqnarray}}
\newcommand{\eea}{\end{eqnarray}}
\newcommand{\bc}{\begin{center}}
\newcommand{\ec}{\end{center}}

\def\gsim{ \lower .75ex \hbox{$\sim$} \llap{\raise .27ex \hbox{$>$}} }
\def\lsim{ \lower .75ex \hbox{$\sim$} \llap{\raise .27ex \hbox{$<$}} }

\usepackage{afterpage}

\usepackage{amssymb}

\usepackage{times}

\usepackage[bookmarks,bookmarksnumbered,colorlinks=true,citecolor=blue,linkcolor=black]{hyperref}
\usepackage{color}

\usepackage{pbox}

\usepackage[flushleft]{threeparttable}

\definecolor{darkgreen}{rgb}{0.0, 0.5, 0.0}

\newcommand{\Msun}{{\rm M_{\odot}}}

\renewcommand{\thefootnote}{\fnsymbol{footnote}}

\title[Galaxy merger rates in Illustris]{The merger rate of galaxies in the Illustris Simulation: \\ a comparison with observations and semi-empirical models}
	
\author[V. Rodriguez-Gomez et al.]
{
	\parbox{18cm}{
		Vicente Rodriguez-Gomez$^{1 \star}$,
		Shy Genel,$^{1,2 \dagger}$
		Mark Vogelsberger,$^{3}$
		Debora Sijacki,$^{4}$ \\
		Annalisa Pillepich,$^{1}$
		Laura V. Sales,$^{1,5}$
		Paul Torrey,$^{3,6}$
		Greg Snyder,$^{7}$
		Dylan Nelson,$^{1}$ \\
		Volker Springel,$^{8,9}$
		Chung-Pei Ma,$^{10}$
		and Lars Hernquist$^{1}$
	}
	\vspace{0.3cm} \\ 
	$^1$ Harvard-Smithsonian Center for Astrophysics, 60 Garden Street, Cambridge, MA 02138, USA \\
	$^2$ Department of Astronomy, Columbia University, 550 West 120th Street, New York, NY 10027, USA \\
	$^3$ Department of Physics, Kavli Institute for Astrophysics and Space Research, Massachusetts Institute of Technology, Cambridge, MA 02139, USA \\
  $^4$ Kavli Institute for Cosmology, Cambridge, and Institute of Astronomy, Madingley Road, Cambridge, CB3 0HA, UK \\
	$^5$ Department of Physics \& Astronomy, University of California, Riverside, 900 University Avenue, Riverside, CA 92521, USA \\
  $^6$ TAPIR, Mailcode 350-17, California Institute of Technology, Pasadena, CA 91125, USA \\
	$^7$ Space Telescope Science Institute, 3700 San Martin Drive, Baltimore, MD 21218, USA \\
	$^8$ Heidelberg Institute for Theoretical Studies, Schloss-Wolfsbrunnenweg 35, 69118 Heidelberg, Germany \\
	$^9$ Zentrum f\"{u}r Astronomie der Universit\"{a}t Heidelberg, ARI, M\"onchhofstr. 12-14, 69120 Heidelberg, Germany \\
	$^{10}$ Department of Astronomy, University of California, Berkeley, CA 94720, USA
}

\begin{document}


\maketitle
\begin{abstract}
We have constructed merger trees for galaxies in the Illustris Simulation by directly tracking the baryonic content of subhalos. These merger trees are used to calculate the galaxy-galaxy merger rate as a function of descendant stellar mass, progenitor stellar mass ratio, and redshift. We demonstrate that the most appropriate definition for the mass ratio of a galaxy-galaxy merger consists in taking both progenitor masses at the time when the secondary progenitor reaches its maximum stellar mass. Additionally, we avoid effects from `orphaned' galaxies by allowing some objects to `skip' a snapshot when finding a descendant, and by only considering mergers which show a well-defined `infall' moment. Adopting these definitions, we obtain well-converged predictions for the galaxy-galaxy merger rate with the following main features, which are qualitatively similar to the halo-halo merger rate except for the last one: a strong correlation with redshift that evolves as $\sim (1+z)^{2.4-2.8}$, a power law with respect to mass ratio, and an increasing dependence on descendant stellar mass, which steepens significantly for descendant stellar masses greater than $\sim 2 \times 10^{11} \, \Msun$. These trends are consistent with observational constraints for medium-sized galaxies ($M_{\ast} \gtrsim 10^{10} \, \Msun$), but in tension with some recent observations of the close pair fraction for massive galaxies ($M_{\ast} \gtrsim 10^{11} \, \Msun$), which report a nearly constant or decreasing evolution with redshift. Finally, we provide a fitting function for the galaxy-galaxy merger rate which is accurate over a wide range of stellar masses, progenitor mass ratios, and redshifts.
\end{abstract}


\section{Introduction}\label{sec:intro}
\renewcommand{\thefootnote}{\fnsymbol{footnote}}
\footnotetext[1]{E-mail: vrodriguez-gomez@cfa.harvard.edu}
\footnotetext[2]{Hubble fellow.}

\renewcommand{\thefootnote}{\arabic{footnote}}

Structure formation in $\Lambda$CDM cosmological models is hierarchical in nature, which makes galaxy mergers an essential aspect of galaxy formation and evolution. In particular, it is important to quantify the galaxy-galaxy merger rate, namely, the frequency of galaxy mergers as a function of the masses of the objects involved, redshift, and possibly other parameters such as gas fractions. A precise determination of this quantity is of fundamental interest for understanding the growth and assembly of galaxies, for bringing galaxy formation models into agreement with the observed distribution of galaxy morphologies, and for explaining the frequency of starburst galaxies and active galactic nuclei at high redshifts.


Although significant progress has been made in the determination of dark matter (DM) halo-halo merger rates using N-body cosmological simulations \citep[e.g.,][]{Fakhouri2008, Genel2009, Genel2010, Fakhouri2010}, with most theoretical predictions agreeing within a factor of $\sim 2$, similar convergence has yet to be achieved in the determination of the galaxy-galaxy merger rate, in particular using theoretical models of galaxy formation and evolution.

There are three main approaches for making theoretical predictions of the galaxy-galaxy merger rate: (1) semi-empirical methods, which typically use an N-body cosmological simulation and `populate' DM subhalos with galaxies according to observational constraints, in particular by applying the halo occupation distribution (HOD) or abundance matching formalisms, (2) semi-analytic models (SAMs), which use an N-body cosmological simulation as the `backbone' of a galaxy formation model, which is implemented in postprocessing, and (3) hydrodynamic simulations, which model the DM and baryonic components of a cosmological volume self-consistently. Therefore, the main difference between SAMs or hydrodynamic simulations with respect to semi-empirical methods is that the latter do not attempt to model galaxy formation processes from first principles (i.e., in an \textit{a priori} fashion), therefore avoiding many of the associated difficulties.

Perhaps the best known determination of the galaxy-galaxy merger rate using a SAM is the one by \cite{Guo2008}, although several other examples can be found in \cite{Hopkins2010a}. On the other hand, there have been very few attempts to determine the galaxy-galaxy merger rate using hydrodynamic simulations \citep[e.g.,][]{Maller2006, Kaviraj2014} due to the fact that until recent years it was not possible to produce statistically significant and sufficiently realistic populations of galaxies in cosmological hydrodynamic simulations.

In general, calculations of the galaxy-galaxy merger rate using semi-empirical methods \citep[][]{Stewart2009, Hopkins2010} are in relatively good agreement with each other, while predictions of the galaxy-galaxy merger rate obtained from various SAMs and hydrodynamic simulations show discrepancies of about an order of magnitude between them, as demonstrated in \cite{Hopkins2010a}. In order to resolve these discrepancies, further work on galaxy merger rates using \textit{a priori} models of galaxy formation is required. This approach has several advantages, such as providing insight into the physical mechanisms included in the models, making predictions in situations where observational data is unavailable, and accounting for merger time-scales self-consistently.

Observational estimates of the galaxy-galaxy merger rate have also not converged yet, although significant progress has been made in this direction \citep{Lotz2011}. For instance, in the case of massive galaxies ($M_{\ast} \gtrsim 10^{11} \, \Msun$), some studies find an increasing redshift dependence \citep{Bundy2009, Bluck2009, Bluck2012, Man2012}, while others find a nearly constant or even decreasing redshift evolution \citep{Williams2011, Newman2012}. Recently, \cite{Man2014} compared the consequences of selecting major mergers by stellar mass and by flux ratio, concluding that the former approach leads to a decreasing redshift dependence, while the latter results in the opposite. This appears to reconcile the differences between the observations by \cite{Bluck2009, Bluck2012} and \cite{Man2012} with those by \cite{Williams2011} and \cite{Newman2012}, where major mergers were selected according to their flux and stellar mass ratios, respectively. However, this is in conflict with the increasing redshift evolution observed for medium-sized galaxies ($M_{\ast} \gtrsim 10^{10} \, \Msun$), as demonstrated in \cite{Lotz2011}, as well as with predictions from semi-empirical models \citep{Stewart2009, Hopkins2010}. The dependence on stellar mass of the galaxy-galaxy merger rate is also a subject of some discussion, with some studies finding an increasing mass dependence and others the opposite \citep[see][for a review]{Casteels2014a}.

We point out that the galaxy-galaxy merger rate cannot be measured directly from observations. Instead, the merger \textit{fraction} must be estimated first, typically from observations of close pairs or morphologically disturbed galaxies, and then converted into a merger rate by adopting some averaged `observability' time-scale \citep{Lotz2011}. Nevertheless, the merger rate and the merger fraction have many common features, such as their evolution with redshift (assuming that the observability time-scales do not change significantly with redshift). For this reason, we will sometimes use the two terms interchangeably when comparing to observations.

In this work we study the galaxy-galaxy merger rate using the Illustris Simulation, a hydrodynamic cosmological simulation carried out in a periodic box of $\sim$106.5 Mpc on a side, which has been shown to reproduce many important properties of galaxies at $z=0$ \citep{Vogelsberger2014a, Vogelsberger2014} as well as at higher redshifts \citep{Genel2014a}. Because of the large volume covered by the simulation, the self-consistent treatment of baryons, and the physically motivated galaxy formation model used \citep{Vogelsberger2013}, the Illustris Simulation provides a unique opportunity to study the galaxy-galaxy merger rate with unprecedented precision and physical fidelity.

This paper is organized as follows. In Section \ref{sec:simulations}, we briefly describe the suite of simulations from the Illustris Project, as well as the methods used to identify halos and galaxies. Section \ref{sec:merger_trees} presents the methodology used to construct merger trees of galaxies and DM halos. The merger rate of DM halos is calculated and compared to previous theoretical work in Section \ref{sec:halo_merger_rate}. We present the definitions and methods used to calculate the galaxy-galaxy merger rate in Section \ref{subsec:merger_rate_definitions}, and in Section \ref{subsec:mass_ratio} we compare different approaches for estimating the mass ratio of a merger. We furthermore explore the dependence of the galaxy-galaxy merger rate as a function of descendant mass, progenitor mass ratio, and redshift in Section \ref{subsec:merger_rate_results}, and compare our results with previous work based on observations and semi-empirical models in Section \ref{subsec:observations}. We finally present a fitting function for the galaxy-galaxy merger rate in Section \ref{subsec:fitting_formula}. We discuss our results and present our conclusions in Section \ref{sec:discussion}.

\section{The simulations}\label{sec:simulations}

  \subsection{Overview}

		The Illustris Project \citep{Vogelsberger2014a, Vogelsberger2014, Genel2014a} is a suite of hydrodynamic cosmological simulations of a periodic box of $75 h^{-1}$ Mpc $\approx$ 106.5 Mpc on a side, carried out with the moving mesh code \textsc{arepo} \citep{Springel2010}. A fiducial physical model has been adopted in these simulations, which includes star formation and evolution, primordial and metal-line cooling with self-shielding corrections, gas recycling and chemical enrichment, stellar supernova feedback, and supermassive black holes with their associated feedback. This model has been described and shown to reproduce several key observables in \cite{Vogelsberger2013}, while its implications for galaxies across different redshifts have been discussed in \cite{Torrey2014}. This model has also been used in hydrodynamic simulations of Milky Way-sized halos \citep{Marinacci2013} and dwarf galaxies \citep{Vogelsberger2014c}.

		The largest simulation from the Illustris project, Illustris-1 (also referred to as \textit{the} Illustris Simulation), follows the dynamical evolution of $2 \times 1820^3$ resolution elements ($1820^3$ DM particles and approximately $1820^3$ gas cells or stellar/wind particles), in addition to $1820^3$ passively evolved Monte Carlo tracer particles. Two lower resolution simulations, Illustris-2 and Illustris-3, follow the dynamical evolution of $2 \times 910^3$ and $2 \times 455^3$ resolution elements, respectively. There are also DM-only variants of the simulations, known as Illustris-Dark-1, Illustris-Dark-2 and Illustris-Dark-3, which can be used to study the effects of baryons on DM halos and subhalos. Each simulation produced 136 snapshots between $z=46$ and $z=0$. The 61 snapshots at $z>3$ are spaced with $\Delta \log_{10}(1+z) \approx 0.02$, while the 75 snapshots at $z<3$ are spaced with $\Delta t \approx 0.15$ Gyr.

The cosmological parameters used throughout this paper are $\Omega_m = 0.2726$, $\Omega_{\Lambda} = 0.7274$, $\Omega_b = 0.0456$, $\sigma_8 = 0.809$, $n_s = 0.963$ and $h = 0.704$, which are consistent with the nine-year \textit{Wilkinson Microwave Anisotropy Probe} (WMAP) measurements \citep{Hinshaw2013}. Unless otherwise noted, all results presented in this paper are derived from Illustris-1.

  \subsection{Identifying the substructure}

		DM halos are identified using the standard friends-of-friends (FoF) approach \citep{Davis1985} with a linking length equal to 0.2 times the mean inter-particle separation. The algorithm is applied to the DM particles, keeping only halos with at least 32 DM particles. After this step, baryonic resolution elements are assigned to the same FoF group as their nearest DM particle. Substructure within the FoF groups is identified using an extension of the \textsc{subfind} algorithm \citep{Springel2001, Dolag2009a}, which can be applied to hydrodynamic simulations.
		
		The original version of \textsc{subfind} \citep{Springel2001} estimates the density field using adaptive kernel interpolation and then identifies subhalo candidates as locally overdense regions. The boundary of each subhalo candidate is determined by the first isodensity contour that passes through a saddle point of the density field. Each subhalo candidate is then subjected to a gravitational unbinding procedure, so that the remaining structures are self-bound. Particles from satellite subhalos which are dropped during the unbinding procedure are tentatively added to the central subhalo (also known as \textit{background} halo) from the same FoF group, which is checked again for gravitational boundness at the end of the process.

		In the version of \textsc{subfind} used with \textsc{arepo}, the density field is calculated for all particles and gas cells using an adaptive smoothing length corresponding to the distribution of DM particles around each point. Subhalo candidates are defined in the same way as before, but during the unbinding procedure the gas thermal energy is also taken into account. We keep subhalos with at least 20 resolution elements (including gas and stars).
		
		We point out that the stellar masses used throughout this paper are the ones given by \textsc{subfind}, \textit{without} truncating the particles found outside a fiducial radius equal to twice the stellar half mass radius \citep{Vogelsberger2014, Genel2014a}. We find that using this alternative definition does not change the galaxy merger rate by more than 10 per cent.

\section{Constructing merger trees}\label{sec:merger_trees}
	
	In this section we describe the algorithms used to construct merger trees. The code for creating subhalo merger trees has been featured in the \textit{Sussing Merger Trees} comparison project \citep{Srisawat2013a, Avila2014, Lee2014}, where it is referred to as \textsc{SubLink}. Essentially, merger trees are constructed at the subhalo level using a methodology similar to the one described in \cite{Springel2005b} and \cite{Boylan-Kolchin2009}, with slight modifications in the merit function used to determine the descendants, a different definition for the \textit{first progenitor} (also known in the literature as the \textit{main progenitor}), and a new method for skipping snapshots. Furthermore, merger trees can be constructed for different particle types, such as DM, stars, and star-forming gas, as explained below.
	
	We define two varieties of merger trees: (1) \textit{DM-only}, which follow exclusively the DM particles of a simulation, and (2) \textit{baryonic}, which follow the star particles \textit{plus} the star-forming gas elements in the simulation. A gas cell is considered to be star-forming if its hydrogen particle density is above 0.13 cm$^{-3}$ \citep{Springel2003}. We note that although following a gas cell is not entirely equivalent to following a collisionless stellar or DM particle, the hydrodynamic scheme implemented in \textsc{arepo} is quasi-Lagrangian, which means that the cells of the moving mesh follow the gas flow to a large extent. Therefore, we assume that star-forming gas cells, which are typically found in the central, denser regions of subhalos, are able to preserve their `identity' for durations of at least a few snapshots, and can therefore add valuable information when determining the descendant of a given subhalo. A less approximate treatment is in principle possible -- although not done here -- by following Monte Carlo tracer particles instead of gas cells \citep{Nelson2013b, Genel2013}. We find that including star-forming gas besides only stellar particles is very useful for constructing robust merger trees at high redshifts, where galaxies have relatively large gas contents.

	If a subhalo does not contain any stars or star-forming gas, then it does not exist in the baryonic merger trees. Conversely, a subhalo without any DM particles does not exist in the DM-only trees (although this situation is extremely rare). The DM-only and baryonic merger trees of the Illustris-1 Simulation contain approximately $5 \times 10^8$ and $7 \times 10^7$ objects, respectively, taking all 136 snapshots into account. All results in this paper were obtained using the baryonic merger trees, with the exception of Section \ref{sec:halo_merger_rate}, where we present results about the merger rate of DM halos rather than galaxies.

	\subsection{Finding the descendants}\label{subsec:descendants}

		Each subhalo is assigned a \textit{unique} descendant (if any) from the next snapshot, an approximation which is consistent with the hierarchical buildup of structure in $\Lambda$CDM cosmologies. This is done in three steps. First, descendant candidates are identified for each subhalo as those subhalos in the following snapshot that have common particles with the subhalo in question. Second, each descendant candidate is given a score based on the following merit function:		
		\begin{equation}
			\chi = \sum_j {\cal R}_j^{-1},	
			\label{eq:merit1}
		\end{equation}
		where ${\cal R}_j$ denotes the binding energy rank of particles from the subhalo in question which are also contained in the descendant candidate. In the case of the baryonic merger trees, equation (\ref{eq:merit1}) is modified to include the mass $m_j$ (taken at the same time as the binding energy rank ${\cal R}_j$) of the resolution elements:
		\begin{equation}
			\chi = \sum_j m_j {\cal R}_j^{-1}.
			\label{eq:merit2}
		\end{equation}
		Third, the unique descendant of the subhalo in question is defined as the descendant candidate with the highest score.
		
		It is worth mentioning that the merit function presented in \cite{Boylan-Kolchin2009} features an exponent of $-2/3$ instead of $-1$. We find that an exponent of $-1$ allows the algorithm to follow subhalos more robustly in major merger scenarios, particularly when three or more objects of comparable sizes and densities interact. Since the outer regions of subhalos are subject to numerical truncation at the saddle points of the density field, as well as physical stripping, one should prioritize tracking the central parts of subhalos, which are the ones that survive the longest. We find that the central few particles of a subhalo are remarkably stable over long periods of time. This is in agreement with previous work \citep[][]{Springel2001, Wetzel2009a}, in which reliable merger trees have been constructed by tracking only the 10--20 most bound particles of each subhalo.

	\begin{figure}
		\centering
	  \includegraphics[width=6.5cm]{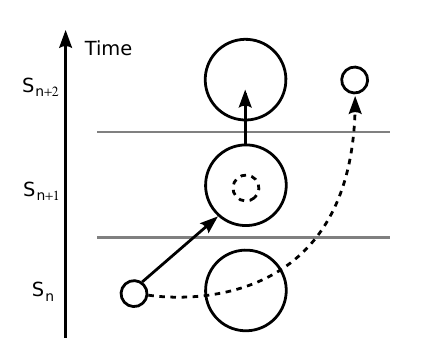}
	  \caption{Illustration of snapshot `skipping,' a simple approach for handling flyby events. The arrows indicate descendance links. A small subhalo identified at snapshot $S_{n}$ is `lost' during snapshot $S_{n+1}$ because it is passing through a larger, denser object. In order to keep track of subhalos in situations like this one, a descendant is also determined at snapshot $S_{n+2}$. If the `skipped' descendant (dashed arrow) is not the same object as the `descendant of the descendant' (solid arrows), then we define the `skipped' one as the correct, unique descendant.}
	  \label{fig:skip_snapshot}
	\end{figure}

		Sometimes a halo finder cannot detect a small subhalo that is passing through a larger structure, simply because the density contrast is not high enough (see Figure \ref{fig:skip_snapshot}). We address this issue in the following way. For each subhalo from snapshot $S_n$, a `skipped descendant' is identified at $S_{n+2}$, which is then compared to the `descendant of the descendant' at the same snapshot. If the two possible descendants at $S_{n+2}$ are not the same object, we keep the one obtained by skipping a snapshot since, by definition, it is the one with the largest score at $S_{n+2}$. This allows us to deal with flyby events, as long as the smaller subhalo is not `lost' during more than one snapshot.
		
		The similarity between the different merger tree algorithms compared in \cite{Srisawat2013a} and \cite{Lee2014} suggests that allowing the search for descendants to extend over exactly two snapshots is enough for most cosmological simulations, which have relatively coarse snapshot spacings. However, a cosmological simulation with extremely high time resolution may require extending the search for descendants over more than two snapshots. This will be explored in future work using the small subboxes described in \cite{Vogelsberger2014}, which have 3976 snapshots each.
		
		The validity of the single-descendant assumption can be investigated by quantifying `how hard' it is to select the best descendant candidate. We did this in the following way. For each galaxy, we calculated the ratio between the `scores' of the best and second-best descendant candidates, namely, $\xi$ = score(second)/score(first). We found that $\xi > 0.5$ ($\xi > 0.1$) in 1 per cent (4 per cent) of the cases. This indicates that, although the approximation is certainly not perfect, in most cases the decision is an `easy' one.

	\subsection{Merger trees of subhalos and galaxies}\label{subsec:merger_trees}

		We say that subhalo $A$ is a progenitor (sometimes also called `direct' progenitor, to distinguish it from earlier progenitors) of subhalo $B$ if and only if subhalo $B$ is the descendant of subhalo $A$. Note that a subhalo can have many progenitors, but at most a single descendant, an approximation motivated by the hierarchical buildup of structure in the Universe.

		Once all the descendant connections have been made, as described in Section \ref{subsec:descendants}, the first progenitor of each subhalo is defined as the one with the `most massive history' behind it \citep{DeLucia2007}. This removes the arbitrariness in defining the first progenitor as simply the most massive one, which is subject to noise when the two largest progenitors have similar masses. As a result, the mass history of any particular galaxy or halo can be robustly compared across simulations carried out at different numerical resolutions or with variations in the physical model, as long as the initial conditions are the same.

		Knowledge of all the subhalo descendants, along with the definition of the first progenitor, uniquely determines the merger trees. However, it is often convenient to rearrange this information into a more useful and physically motivated form. For example, one might be interested in retrieving the mass of a given object for all previous times, which would be a burdensome task if given the raw descendant information alone. We therefore construct merger trees in the following way. First, a linked-list structure is created for the whole simulation, so that each subhalo is assigned pointers to five `key' subhalos \citep{Springel2005b}:
		
		\begin{itemize}
			\item[] \textit{First progenitor:} The progenitor of the subhalo in question, if any, which has the `most massive history' behind it.
			\item[] \textit{Next progenitor:} The subhalo, if any, which shares the same descendant as the subhalo in question, and which has the next largest `mass history' behind it.
			\item[] \textit{Descendant:} The unique descendant of the subhalo in question, if any.
			\item[] \textit{First subhalo in FoF group:} The main subhalo (defined as the one with the `most massive history' behind it) from the same FoF group as the subhalo in question. Note that this link can point back to the subhalo under consideration.
			\item[] \textit{Next subhalo in FoF group:} The next subhalo from the same FoF group, if any, in order of decreasing `mass history.'
		\end{itemize}

After this, the linked-list structure is stored in a depth-first fashion \citep{Lemson2006} into several files on a `per tree' basis, where each tree is defined as a set of subhalos that are connected by progenitor/descendant links or by belonging to the same FoF group. More specifically, two subhalos belong to the same tree if and only if they can be reached by successively following the pointers described above. The resulting trees are completely independent from each other, which allows for easy parallelization of computationally expensive postprocessing tasks, such as the construction of \textit{halo} merger trees.


%


	\subsection{Merger trees of halos (FoF groups)}\label{subsec:halo_merger_trees}

	\begin{figure*}\centerline{
	\hbox{
	\includegraphics[width=\columnwidth]{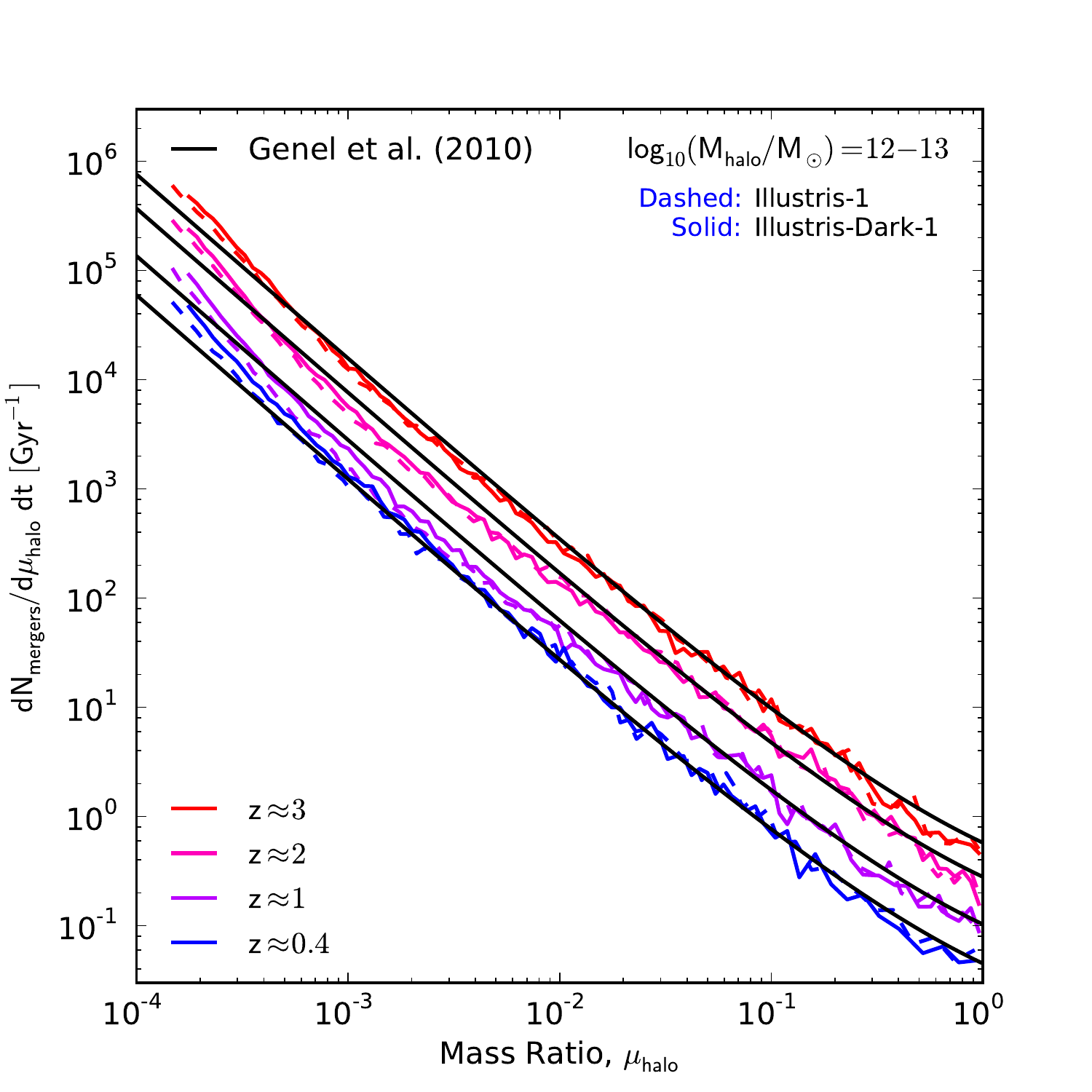}
	\includegraphics[width=\columnwidth]{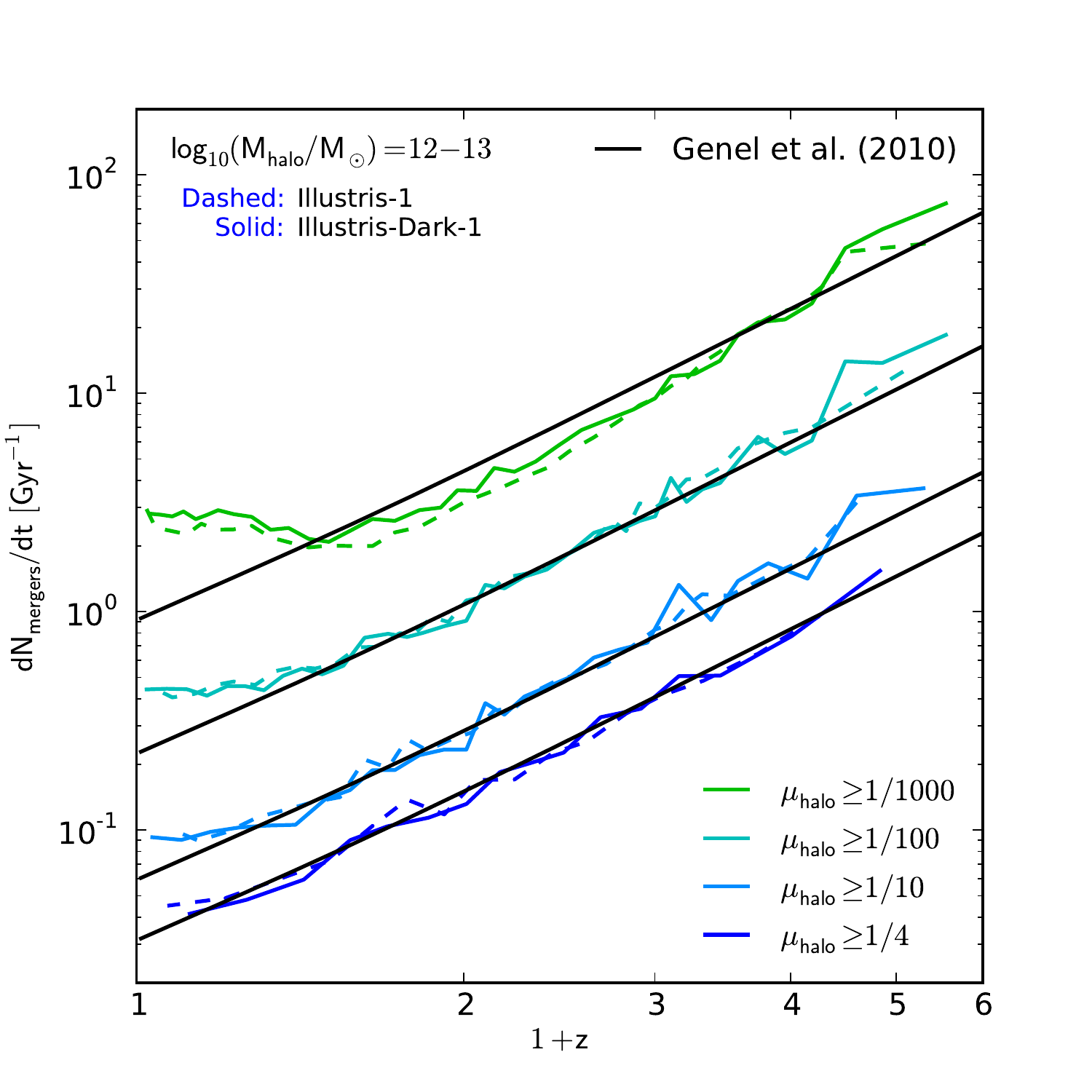}
	}}
	\caption{\textit{Left:} The halo-halo merger rate as a function of the mass ratio $\mu_{\rm halo}$, shown for different redshifts. \textit{Right:} The \textit{cumulative} (with respect to mass ratio) halo-halo merger rate as a function of redshift, shown for different minimum mass ratios. Both panels correspond to mergers with descendant halo masses in the range $10^{12} \leq M_{\rm halo}/\Msun < 10^{13}$. The solid black lines are predictions from the fitting function given in \protect\cite{Genel2010}. The colored dashed and solid lines correspond to the Illustris-1 (baryonic) and Illustris-Dark-1 (DM-only) simulations, respectively. The very good agreement between the dashed and solid lines indicates that baryons do not have a significant influence on the halo-halo merger rate. The increase in the merger rate seen at low redshifts is due to a limitation of the splitting algorithm as it approaches the final snapshots of the simulation, since spurious mergers can only be distinguished from real ones when there is a sufficient number of `future' snapshots.}
	\label{fig:halo_merger_rate}
	\end{figure*}

		Although most of the results in this paper were obtained using \textit{galaxy} merger trees, we also construct halo (i.e., FoF group) merger trees in order to calculate the halo-halo merger rate. This quantity is relatively well constrained by theoretical models, so it can be used to validate some of our most basic results, as well as to assess the effects of cosmic variance on the cosmological volume used for this study, as discussed in Section \ref{sec:halo_merger_rate}.
		
		Halo merger trees can contain fragmentation events in which a halo is split into two (or more) descendant halos. These events arise because particles in a progenitor halo rarely end up in exactly one descendant halo; a decision therefore must be made to select a unique descendant halo. There is not a unique way to do this, and various algorithms have been proposed \citep[see, e.g.,][for a detailed comparison]{Fakhouri2009a}.
		
		Here we construct halo merger trees using the \textit{splitting} algorithm \citep{Genel2009, Genel2010, Fakhouri2009a, Fakhouri2010a}. Instead of tracking the particles from each FoF group directly, this method takes the \textit{subhalo} merger trees as input and constructs \textit{halo} merger trees which are completely free of halo fragmentation events, as described below. The mass of each halo is defined as its \textit{bound} mass, i.e., the combined mass of all particles gravitationally bound to its subhalos, instead of the FoF group mass, which can contain a significant contribution from unbound particles.
		
		Halo fragmentations are removed in the following way. For every tentative merger event between two halos, the splitting algorithm checks whether the two halos separate at a later time (as would happen in the case of a flyby), and, if that is the case, it then considers the two halos as separate objects for all times. More specifically, for every halo at redshift $z_{\rm high}$, the algorithm checks whether the halo contains at least one pair of subhalos which at some lower redshift $z_{\rm low}$ do not belong to the same halo. Such halo would then be split in the following way: two subhalos which belong to different halos at $z_{\rm low}$ will also belong to different halos at $z_{\rm high}$, while subhalos that stay together at $z_{\rm low}$ will also be together at $z_{\rm high}$.
		
		The splitting algorithm yields a new population of DM halos and associated merger trees which are completely free from fragmentations, while leaving the DM halo mass function relatively unchanged \citep{Genel2009}. 

\section{The halo-halo merger rate}\label{sec:halo_merger_rate}
	
	The merger rate of DM halos has been studied extensively in previous work \citep[e.g.,][and references therein]{Fakhouri2008, Genel2009, Genel2010, Fakhouri2010}, with different theoretical predictions being similar within a factor of $\sim$2. Therefore, before calculating the galaxy-galaxy merger rate, we first verify that the halo-halo merger rate in Illustris is consistent with previous work.

	Using the \textit{splitting} method \citep{Genel2009, Genel2010, Fakhouri2009a, Fakhouri2010a}, \textit{halo} merger trees were constructed by taking the DM-only \textit{subhalo} merger trees as input (see Section \ref{subsec:halo_merger_trees}). The resulting halo-halo merger rate is plotted in Figure \ref{fig:halo_merger_rate}, both as a function of mass ratio for different redshifts (left) and as a function of redshift for different minimum mass ratios (right), for halos with total \textit{bound} masses (see Section \ref{subsec:halo_merger_trees}) between $10^{12} \, \Msun$ and $10^{13} \, \Msun$.\footnote{The current implementation of the splitting algorithm supports a single particle type with a fixed mass. For this reason, the masses used to calculate the merger rate in Illustris-1 (dashed lines) actually correspond to the DM components rather than the total masses, which makes them smaller than their Illustris-Dark-1 counterparts (solid lines) by $\sim$20 per cent (without taking baryonic effects into account). However, this difference is negligible for our purposes because of the weak mass dependence of the halo merger rate, $\sim M_{\rm halo}^{0.15}$, which results in a change in the merger rate below 3 per cent. For comparison, the typical errorbar size in both panels of Figure \ref{fig:halo_merger_rate}, produced by Poisson noise in the number of mergers, is $\sim$10--20 per cent. Thus, Figure \ref{fig:halo_merger_rate} would be essentially unchanged if we had used the total mass instead of the DM mass for Illustris-1 halos.} The solid black lines correspond to predictions from the fitting function provided by \cite{Genel2010}. The colored dashed and solid lines show the halo merger rate in the Illustris-1 (baryonic) and Illustris-Dark-1 (DM-only) simulations, respectively. 

The very good agreement between the baryonic and DM-only Illustris runs in Figure \ref{fig:halo_merger_rate} indicates that baryons do not play an important role in the merger rate of halos. Although not shown in this work, we have also calculated the halo merger rate for all the different feedback implementations described in \protect\cite{Vogelsberger2013}, as well as for the \textsc{gadget} and \textsc{arepo} runs described in \protect\cite{Vogelsberger2012}, which also resulted in no significant difference between any of them. This again shows that the halo merger rate is remarkably robust to different implementations of baryonic physics.

	Figure \ref{fig:halo_merger_rate} shows that the halo-halo merger rate in the Illustris Simulation is in excellent agreement with the formula provided by \cite{Genel2010} (except for redshifts $z \lesssim 0.4$, as discussed below). This is noteworthy given the fact that the best-fitting parameter values were obtained using the Millennium and Millennium II Simulations \citep{Springel2005b, Boylan-Kolchin2009}, which were carried out with cosmological parameters different from those in Illustris, and for which different subhalo merger trees were used as input for the splitting algorithm. In agreement with \cite{Genel2010}, we find that the halo-halo merger rate scales with redshift as $\sim (1+z)^{2.3}$, with mass ratio as $\sim \mu_{\rm halo}^{-1.7}$, and with descendant mass as $\sim M_{\rm halo}^{0.15}$. These values are similar to the ones found by \cite{Fakhouri2010}.
	
	The good agreement between the halo merger rate in Illustris and the fit from \cite{Genel2010} also suggests that cosmic variance can be neglected in the 106.5 Mpc box used for this study, i.e., that the initial conditions used in the simulation are indeed representative of the large-scale density field. A detailed discussion about cosmic variance and the choice of initial conditions in Illustris can be found in \cite{Genel2014a}.

	The increase in the merger rate seen at low redshifts is an unavoidable limitation of the splitting algorithm as it approaches the end of the simulation, since it becomes impossible to determine whether two recently merged halos will `remain' merged after $z=0$, and therefore spurious mergers cannot be removed. For this reason, the calculated merger rate at $z \lesssim 0.4$ is overestimated and an extrapolation should be used instead. It is worth mentioning that analytic estimates of the halo merger rate based on the Extended Press-Schechter formalism \citep{Press1974, Bond1991, Lacey1993, Neistein2008a, Neistein2008} predict that the halo merger rate remains roughly a power law with respect to $(1+z)$ up to (and beyond) $z=0$, which justifies the extrapolation used by \cite{Genel2009, Genel2010}.

	Finally, we point out that since we used the splitting algorithm to construct the halo merger trees (Section \ref{subsec:halo_merger_trees}), the fitting formula from \cite{Genel2010} is the only analytical expression that can provide a meaningful comparison with previous work. If we had instead constructed halo merger trees using the \textit{stitching} method \citep{Fakhouri2008}, then the fit from \cite{Fakhouri2010} would be a better description of the resulting data. The halo merger rates obtained using these two methods can differ by up to a factor of 2 at $z \approx 0.4$ \citep[][figures 5 and 6]{Genel2009}.

\section{The galaxy-galaxy merger rate}\label{sec:galaxy_merger_rate}

	In this section we describe how the galaxy-galaxy merger rate was calculated and explore its scaling as a function of descendant stellar mass, progenitor stellar mass ratio, and redshift. We also compare the merger rate with observations from the literature and provide a fitting formula which is reasonably accurate over a large range of masses, mass ratios, and redshifts.

	We point out that the results about galaxy merger rates presented in this section were obtained directly from the \textit{galaxy} merger trees (Section \ref{subsec:merger_trees}). Thus, they are independent from details about \textit{halo} merger trees and rates (Sections \ref{subsec:halo_merger_trees} and \ref{sec:halo_merger_rate}).

	\subsection{Definitions}\label{subsec:merger_rate_definitions}

		\subsubsection{Merger}

			A merger takes place when a galaxy has more than one direct progenitor. Direct progenitors are usually found within the previous snapshot, but in some rare cases they are found two snapshots before, as discussed in Section \ref{sec:merger_trees}.
			
			We assume that all mergers are binary, which means that if a galaxy has $N_p$ direct progenitors, we count $N_p-1$ mergers, each between the first progenitor and each of the other ones. \cite{Fakhouri2008} studied the effects produced by assuming binary versus multiple mergers and determined that for a low-redshift snapshot spacing of $\Delta z = 0.02$, the binary counting method was a good approximation for a wide range of halo masses and mass ratios. The low-redshift snapshot spacing in Illustris is $\Delta z \approx 0.01$, i.e., two times smaller than the value recommended by \cite{Fakhouri2008}, which means that Illustris is in the `safe' regime with respect to the binary counting approximation.
			
			Each merger is characterized by three parameters:
			\begin{itemize}
				\item[] $M_{\ast}$: The stellar mass of the descendant immediately \textit{after} the merger takes place.\footnote{Although we could include star-forming gas in the mass of a galaxy, as we did when constructing the baryonic merger trees, for the rest of this paper we shall mostly be concerned with the stellar mass, since this quantity can be more directly compared to observations. This is consistent with our goal of quantifying the frequency of mergers, rather than the effects produced by them (in which case the gas content would indeed play an important role).}
				\item[] $\mu_{\ast}$: The ratio between the stellar masses of the primary and secondary progenitors, taking both masses at $t_{\rm max}$, defined as the moment when the secondary reaches its maximum stellar mass (see Section \ref{subsec:mass_ratio}).
				\item[] $z$: The redshift of the descendant snapshot.
			\end{itemize}

			Most halo finders have difficulty in correctly identifying subhalos (or galaxies) during the final stages of a merger, which leads to `orphaned' subhalos during the construction of merger trees and a subsequent overestimation of the merger rate, since some merger events would be counted more than once. In order to avoid this, we only consider mergers which show a clear \textit{infall} moment, that is, mergers for which both progenitors, followed back in time through their main branches in the merger trees, belonged to different FoF groups at some point in the past. This condition also becomes necessary in connection with the different definitions for the progenitor mass ratio discussed in Section \ref{subsec:mass_ratio}.

		\subsubsection{Merger rate}		

		\begin{figure*}\centerline{\hbox{
		\includegraphics[width=20cm]{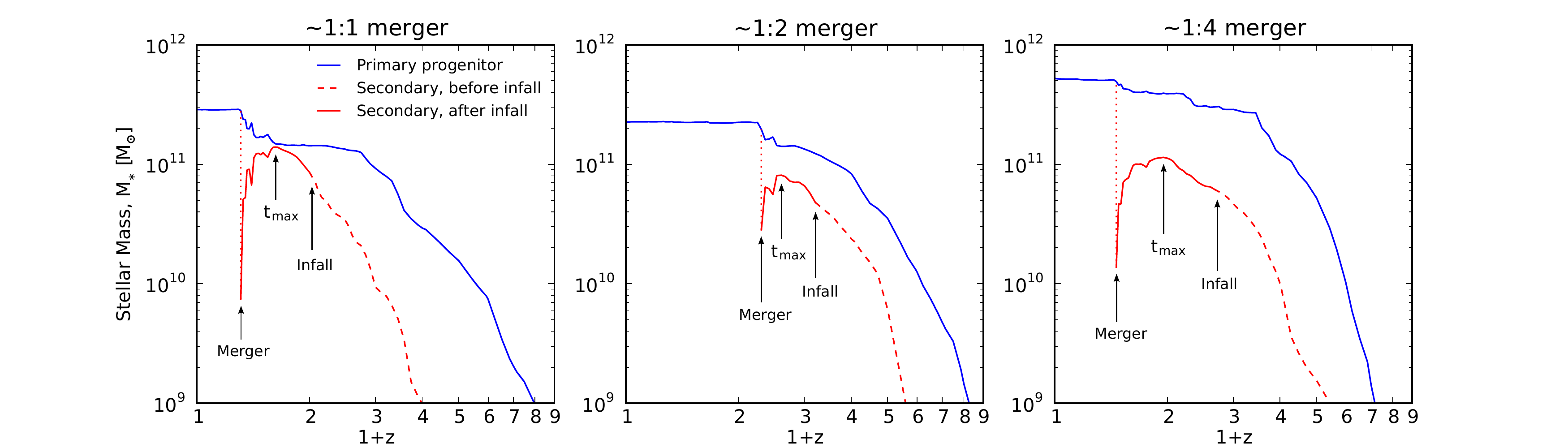}}}
		\caption{Stellar mass as a function of redshift, shown for galaxies undergoing mergers of different mass ratios (approximately 1:1, 1:2 and 1:4, from left to right). In each panel, the blue line corresponds to the main branch of a galaxy identified at $z=0$, while the red line represents the main branch of a secondary galaxy that merges with the primary. The moment when the two galaxies merge is indicated with a vertical dotted line connecting the two progenitors. The secondary progenitor is drawn with a solid line when it is found inside the same FoF group as the primary, and with a dashed one when it is outside. In order to calculate the mass ratio of a merger, the masses of both progenitors are taken at $t_{\rm max}$, i.e., at the time when the \textit{secondary} progenitor reaches its maximum stellar mass. Note that the mass ratio would be severely underestimated if the progenitor masses were taken right before the merger.}
		\label{fig:merger_examples}
		\end{figure*}

			The galaxy-galaxy merger rate describes the frequency of galaxy mergers as a function of descendant stellar mass $M_{\ast}$, progenitor stellar mass ratio $\mu_{\ast}$, and redshift $z$. In this work we focus on the merger rate \textit{per galaxy}, which corresponds to the number of mergers per descendant galaxy, per unit time, per unit mass ratio. This quantity is typically given in units of Gyr$^{-1}$, and we denote it by

		\begin{equation}
		  \frac{{\rm d}N_{\rm mergers}}{{\rm d}\mu_{\ast} \, {\rm d}t} (M_{\ast}, \mu_{\ast}, z).
			\label{eq:merger_rate}
		\end{equation}		

		In practice, Equation (\ref{eq:merger_rate}) can be approximated in four steps: $(i)$ defining bins in $M_{\ast}$, $\mu_{\ast}$ and $z$, $(ii)$ counting the number of mergers that fall into each bin, $(iii)$ dividing by the average number of galaxies per snapshot for each corresponding bin, and $(iv)$ dividing by the time interval, which is determined by the time difference between the snapshots that are located just before the edges of each redshift bin. Note that each redshift bin can contain more than one snapshot.

			We make sure that each bin contains a minimum number of mergers (usually 5 or 10), so that bins are joined together when this is not the case. Additionally, we impose a resolution limit of at least 10 stellar particles for the smallest progenitor in each merger. The uncertainty in the calculated merger rate is determined by the Poisson noise from the number of mergers in each bin.\footnote{In general, there are many more galaxies than mergers for any of the time-scales considered, so we neglect the error contribution from the number of galaxies in each bin.}

			Finally, since we are defining the first progenitor as the one with the `most massive history' behind it, rather than as simply the most massive one, it is possible to have mass ratios greater than one. In these cases we invert the mass ratio, so that we always have $\mu_{\ast} \leq 1$. We find that this minor correction has a negligible effect for all our results, with the resulting merger rate being indistinguishable from the one obtained by simply discarding mergers with $\mu_{\ast} > 1$ (the difference is much smaller than the uncertainty produced by the Poisson noise from the number of mergers).

		\subsection{The mass ratio of a merger}\label{subsec:mass_ratio}

		As mentioned above, the mass ratio of a merger is based on the stellar masses of the two progenitors taken at $t_{\rm max}$, i.e., at the time when the secondary progenitor reaches its maximum stellar mass. Here we provide justification for this choice and explore other alternatives, such as taking the progenitor masses right before the merger and at virial infall.

	\begin{figure*}\centerline{
	\hbox{
	\includegraphics[width=\columnwidth]{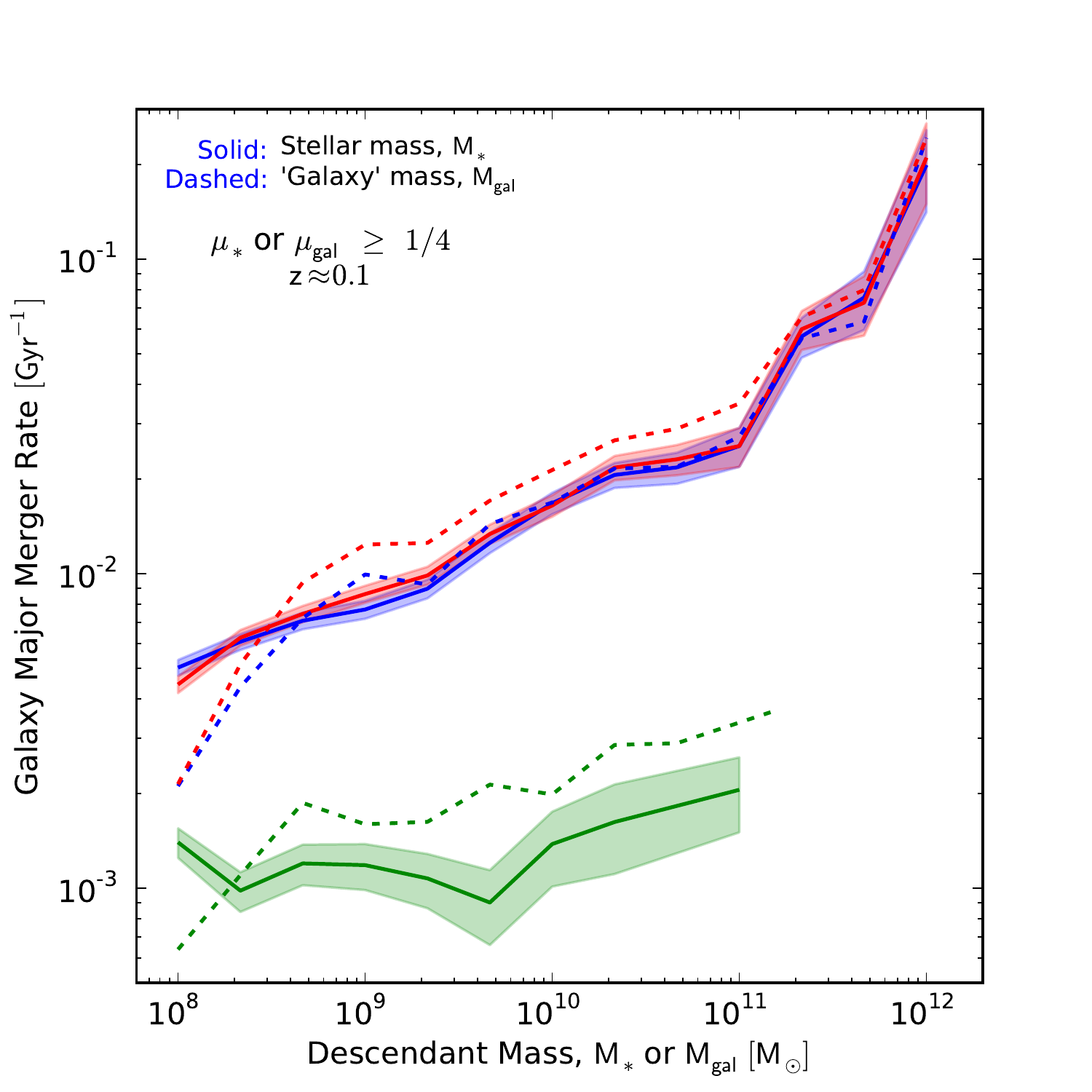}
	\includegraphics[width=\columnwidth]{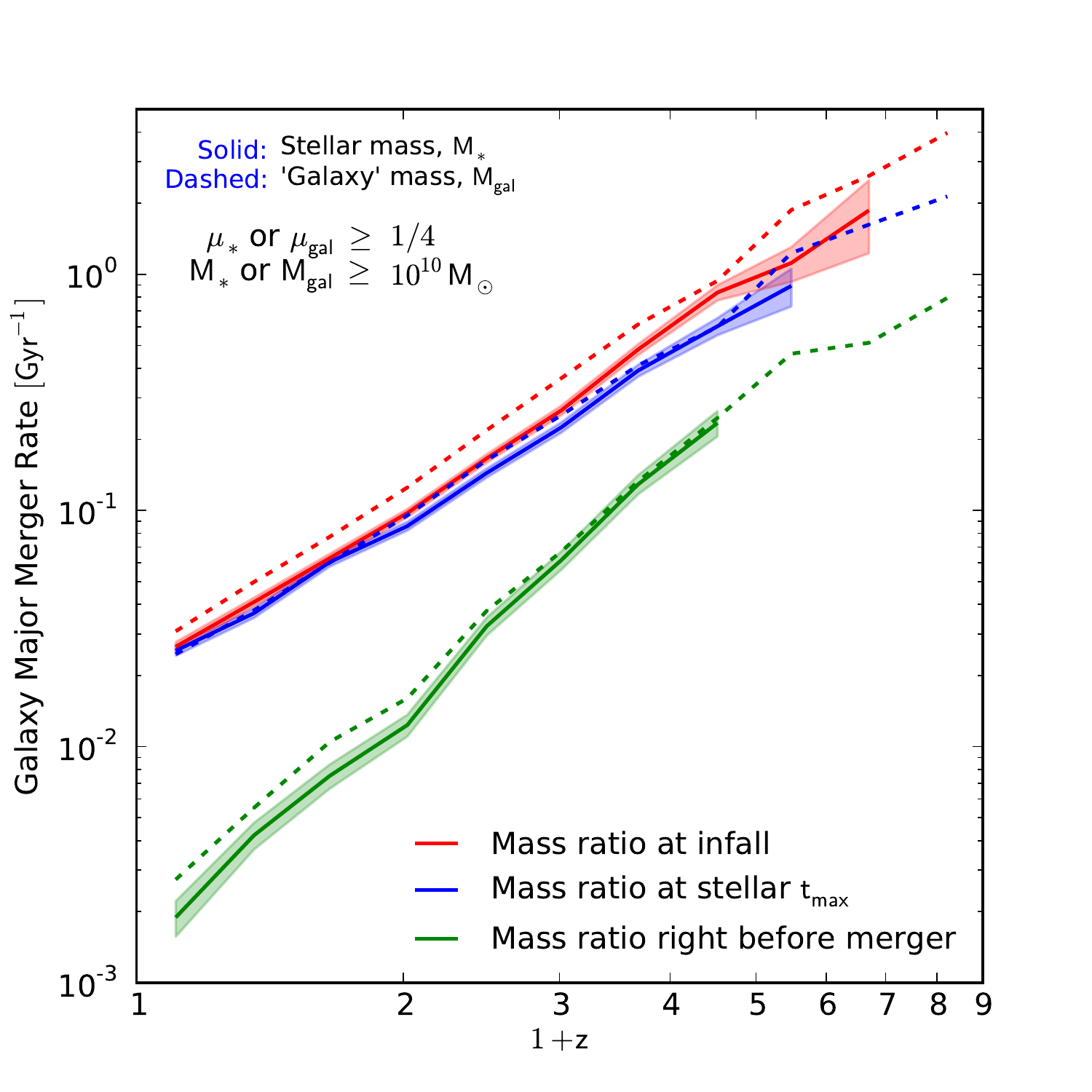}
	}}
	\caption{\textit{Left:} Major merger rate per galaxy as a function of descendant mass, for a redshift bin centered around $z = 0.1$. \textit{Right:} Major merger rate per galaxy as a function of redshift, for descendant galaxy masses greater than $10^{10} \, \Msun$. The different colors show merger rates calculated by taking the mass ratio at different times, while the solid and dashed lines indicate merger rates calculated by using stellar and `galaxy' (stars plus star-forming gas) masses, respectively. The shaded regions represent the Poisson noise from the number of mergers in each bin. We observe that taking the progenitor masses right before a merger can severely underestimate the major merger rate.}
	\label{fig:manydefs}
	\end{figure*}

	\begin{figure*}\centerline{
	\hbox{
	\includegraphics[width=\columnwidth]{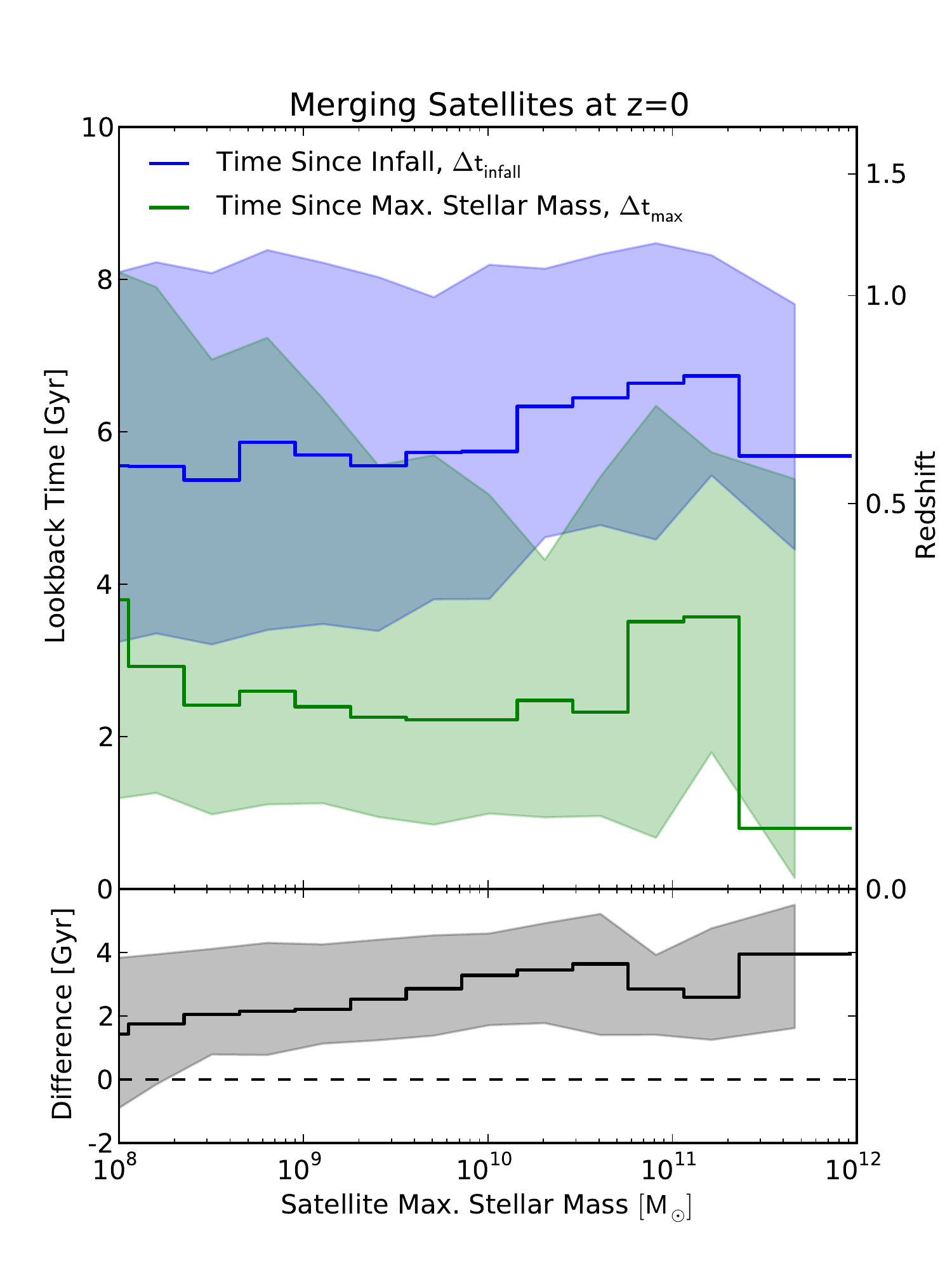}
	\includegraphics[width=\columnwidth]{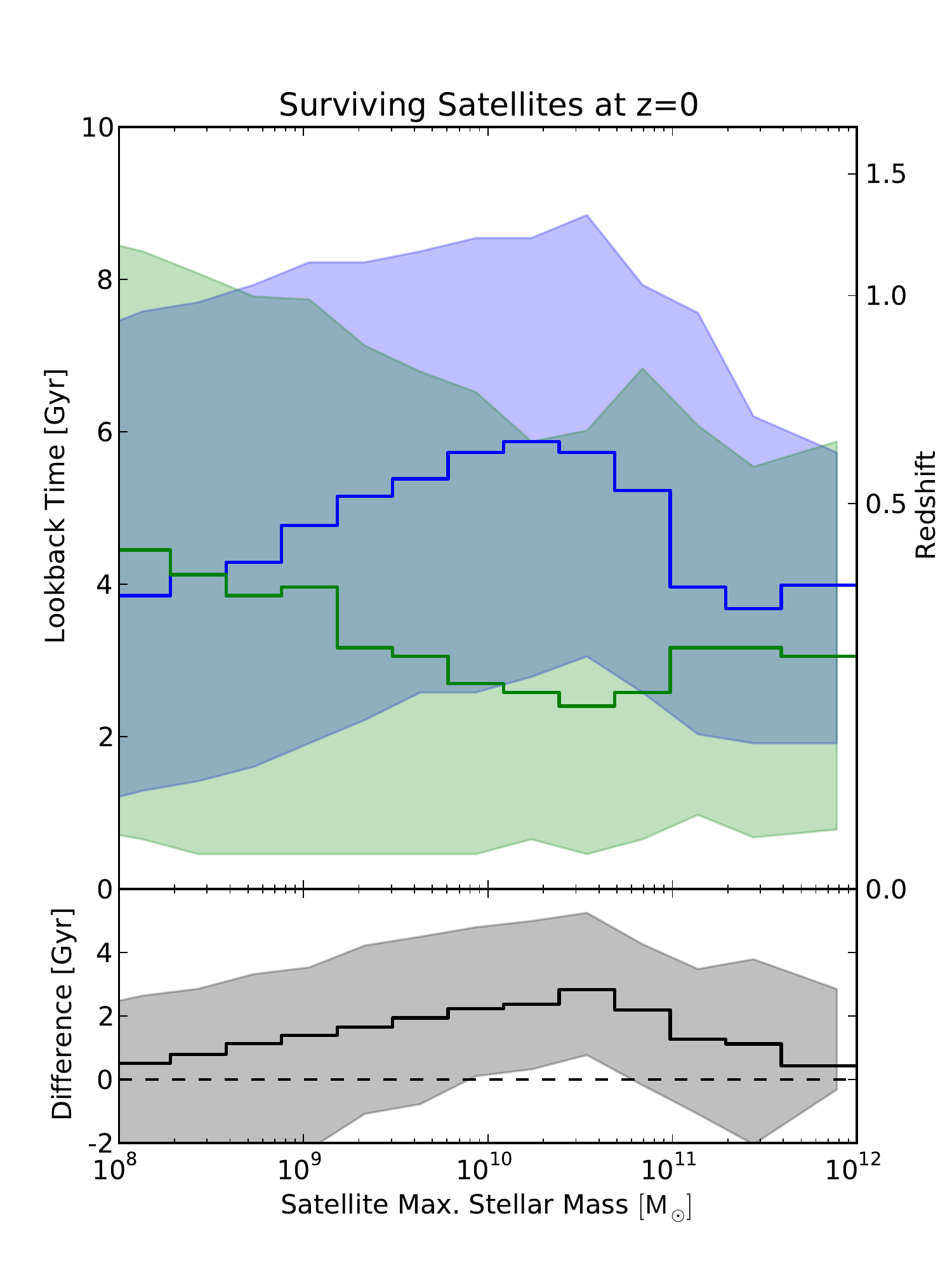}
	}}
	\caption{\textit{Left:} The median elapsed times since virial infall ($\Delta t_{\rm infall}$, blue) and since the moment of maximum stellar mass ($\Delta t_{\rm max}$, green), shown for \textit{merging} satellites at $z=0$ as a function of maximum stellar mass. \textit{Right:} The same for \textit{surviving} satellites at $z=0$. The bottom panels show the median of the difference between $t_{\rm infall}$ and $t_{\rm max}$, calculated for each galaxy. The shaded regions indicate the range between the 16th and 84th percentiles, or approximately $1 \sigma$ (note that the two shaded regions can overlap, which results in a darker color). We observe that most satellites reach their maximum stellar mass a few Gyr \textit{after} infall. The apparent sign reversal in the right panels around $10^8 \, \Msun$ happens simply because median values are not additive, so that $\rm median(t_{\rm infall}) < median(t_{\rm max})$ does not necessarily imply that $\rm median(t_{\rm infall} - t_{\rm max}) < 0$, and vice versa.}
	\label{fig:mass_timescales}
	\end{figure*}

		Figure \ref{fig:merger_examples} shows typical mass histories of galaxies that are undergoing mergers of different mass ratios, approximately 1:1, 1:2 and 1:4. Each panel shows $(i)$ the moment when the two galaxies merge, $(ii)$ $t_{\rm max}$, the time when the secondary progenitor reaches its maximum stellar mass, and $(iii)$  the infall moment, i.e., the time when the secondary progenitor enters the same FoF group as the primary one.

		In all panels we observe that, shortly before the merger takes place, there appears to be an `exchange' of mass between the primary and secondary progenitors. This is a consequence of how the halo finder imposes a distinction between centrals and satellites: even when two merging objects have nearly identical initial masses, one of them will be defined as the central subhalo and the other one as a satellite. Then, by construction, the central subhalo (or \textit{background} halo) will be assigned most of the loosely bound matter residing in the FoF group, while the satellite will be `truncated' by the saddle points in the density field. As a result, the central is typically much more massive than the satellite, even when the particle distribution of the two objects remains approximately symmetrical. This means that the mass ratio of a merger would be severely underestimated if we took the masses of the progenitors right before the merger.

		Such effects are well known in the context of DM-only simulations. In particular, it has been observed that the DM mass of a satellite subhalo artificially correlates with its distance to the center of the halo \citep[e.g.,][]{Sales2007, Wetzel2009a, Muldrew2011}. Here we show, however, that care must also be taken when considering the stellar content of a subhalo, despite the fact that stars are much more concentrated than DM and therefore less susceptible to numerical truncation.
		
		Although phase-space halo finders seem to capture the masses of subhalos more reliably during major mergers \citep[see][for a review]{Avila2014}, there is an additional reason why we avoid taking the progenitor masses immediately before a merger, which is to avoid effects from the merger itself, such as enhanced star formation and physical (as opposed to numerical) stripping, among other possible effects produced by mergers.
		
		There are alternative definitions for the mass ratio of a merger. The two most relevant ones consist of taking the progenitor masses at $t_{\rm infall}$, the time when the secondary progenitor enters the same FoF group as the primary one, and taking them at $t_{\rm max}$, the time when the secondary progenitor reaches its maximum stellar mass.

		Figure \ref{fig:manydefs} shows the major merger rate as a function of descendant mass (left) and as a function of redshift (right), for the three mass ratio definitions mentioned so far, which are indicated with different colors. Additionally, we show with dashed lines the corresponding merger rates obtained by replacing each stellar mass with the corresponding `galaxy' (baryonic) mass, defined as the stellar \textit{plus} star-forming gas mass of each galaxy.

		Clearly, taking the galaxy masses right before a merger can underestimate the major merger rate by an order of magnitude or more. In fact, we find that both $t_{\rm infall}$ and $t_{\rm max}$ result in galaxy merger rates that are very well converged with resolution, while taking the progenitor masses right before a merger yields a merger rate that becomes smaller with increasing resolution. Indeed, as the resolution of a simulation is increased, two merging galaxies can be individually identified for a longer time before they finally merge, which means that they can get closer to each other, leading to a more extreme mass difference. This means that a major merger ($\mu_{\ast} \geq 1/4$) will appear to be a much more minor one by the time the merger actually takes place, which results in an underestimation of the major merger rate.
		
		Another noticeable trend from Figure \ref{fig:manydefs} is that using baryonic masses instead of stellar masses results in slightly larger merger rates. This is a consequence of the decreasing fraction of cold gas as a function of stellar mass. Indeed, if we make the approximation $M_{\rm gal} \propto M_{\ast}^{\alpha}$, where $\alpha < 1$, then a $\mu_{\ast} \gtrsim 1/4$ (major) merger in baryonic mass would correspond to a more minor one in stellar mass, which might not contribute to the major merger rate in this case.

		In general, we observe that the merger rates obtained by taking the progenitor masses at $t_{\rm infall}$ and $t_{\rm max}$ are very similar, which is a consequence of the mass \textit{ratio} being mostly unchanged between $t_{\rm infall}$ and $t_{\rm max}$. However, even when the mass ratio is similar, the \textit{masses} themselves can be very different across these two times, as a consequence of the large amount of star formation that can take place after infall. This can be seen in the three merger examples from Figure \ref{fig:merger_examples}, where the stellar mass grows by approximately a factor of 2 between $t_{\rm infall}$ and $t_{\rm max}$ \citep[see also][for a discussion about star formation in Illustris satellites after infall and their resulting colors]{Sales2014a}. This suggests that taking the progenitor masses at $t_{\rm infall}$ is too early for making any meaningful comparison with observations of galaxy close pairs, which presumably involve observations of galaxies that have already assembled most of their stellar mass.
			
					\begin{table*}
		\centering

		\begin{tabular}{c | c | c}
			\hline
			Definition & $\frac{{\rm d}N_{\rm mergers}}{{\rm d}\mu \, {\rm d}t} (M, \mu, z)$ \\
			\hline
			Units & Gyr$^{-1}$ \\
			\hline
			\parbox{2cm}{Fitting function} &
			\parbox{8cm}{\centering $A(z) \, \left(\frac{M}{10^{10}\Msun}\right)^{\alpha(z)} \left[1 +  \left(\frac{M}{M_0}\right)^{\delta(z)}\right] \mu^{\beta (z) + \gamma \log_{10}\left(\frac{M}{10^{10}\Msun}\right)}$, \\
			                               where \\
			                               $A(z) = A_0 (1+z)^{\eta}$, \\
			                               $\alpha(z) = \alpha_0 (1+z)^{\alpha_1}$, \\
			                               $\beta(z) = \beta_0 (1+z)^{\beta_1}$, \\
			                               $\delta(z) = \delta_0 (1+z)^{\delta_1}$, \\
			                               and $M_0 = 2 \times 10^{11} \, \Msun$ is fixed.} \\
			\hline
			\parbox{2cm}{\centering
				$\log_{10}(A_0/{\rm Gyr^{-1}})$ \\
				$\eta$ \\
				$\alpha_0$ \\
				$\alpha_1$ \\
				$\beta_0$ \\
				$\beta_1$ \\
				$\gamma$ \\
				$\delta_0$ \\
				$\delta_1$} &
			\parbox{4cm}{\flushright
				$-2.2287 \pm 0.0045$ \\
				$2.4644 \pm 0.0128$ \\
				$0.2241 \pm 0.0038$ \\
				$-1.1759 \pm 0.0316$ \\
				$-1.2595 \pm 0.0026$ \\
				$0.0611 \pm 0.0021$ \\
				$-0.0477 \pm 0.0013$ \\
				$0.7668 \pm 0.0202$ \\
				$-0.4695 \pm 0.0440$} \hspace{2cm} \\

			\hline
			$\chi^2_{\rm red}$ & 1.16 \\
			\hline
		\end{tabular}

		\caption{Fitting function and best-fitting parameters for the galaxy-galaxy merger rate (both $M$ and $\mu$ correspond to stellar masses). See Section \ref{subsec:fitting_formula} for details.}
		\label{tab:fitting_formula}
	\end{table*}

\begin{figure*}
	\centerline{
		\vbox{
			\includegraphics[width=\textwidth]{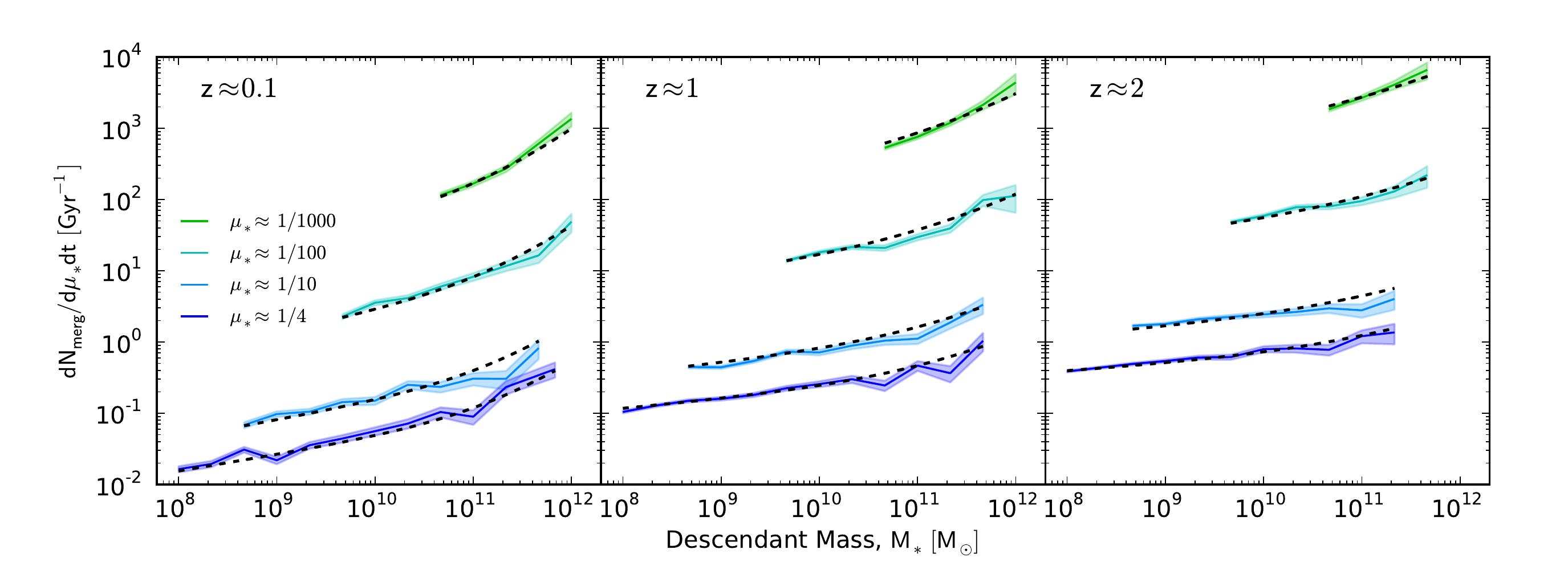}
			\includegraphics[width=\textwidth]{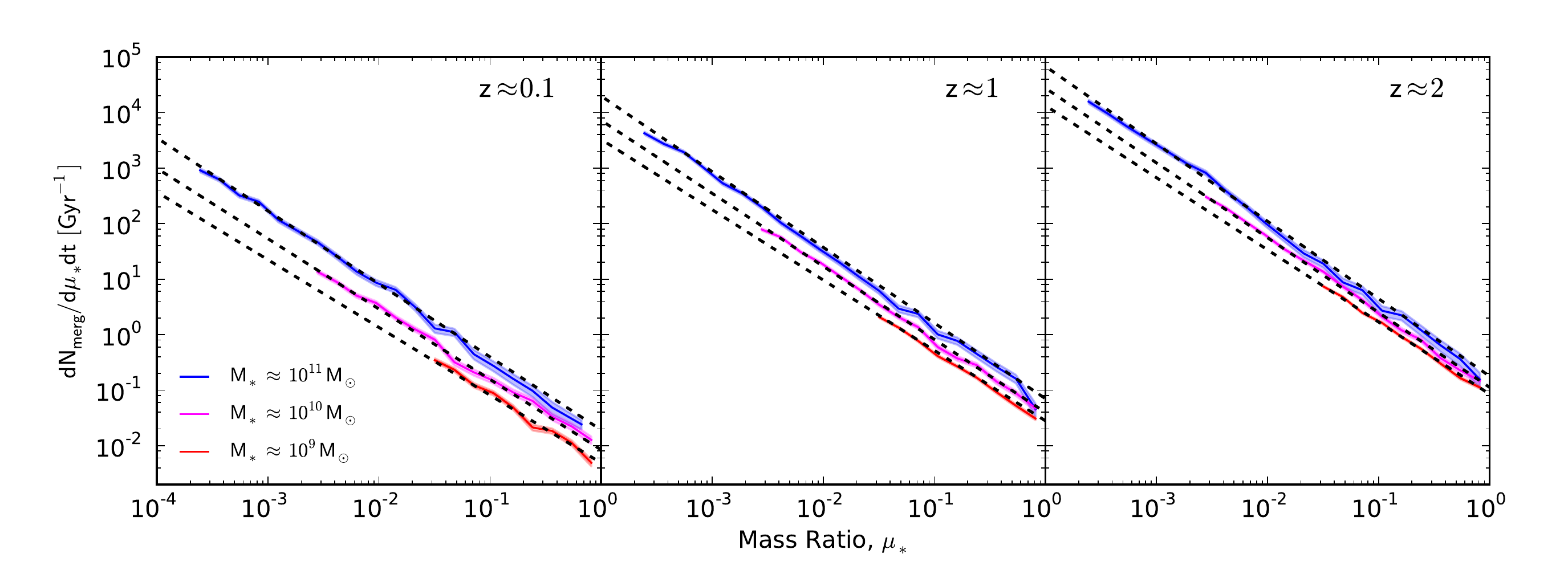}
		}
	}
\caption{\textit{Top:} The galaxy merger rate as a function of descendant mass $M_{\ast}$, for different mass ratios. \textit{Bottom:} The galaxy merger rate as a function of mass ratio $\mu_{\ast}$, for different descendant masses. The left, center, and right panels correspond to redshift bins centered around 0.1, 1, and 2, respectively. The shaded regions indicate the Poisson noise in the number of mergers in each bin. The dashed black line represents the fitting function from Table \ref{tab:fitting_formula}.}
\label{fig:differential_merger_rate}
\end{figure*}

	\begin{figure*}
	\centerline{
		\vbox{
			\includegraphics[width=\textwidth]{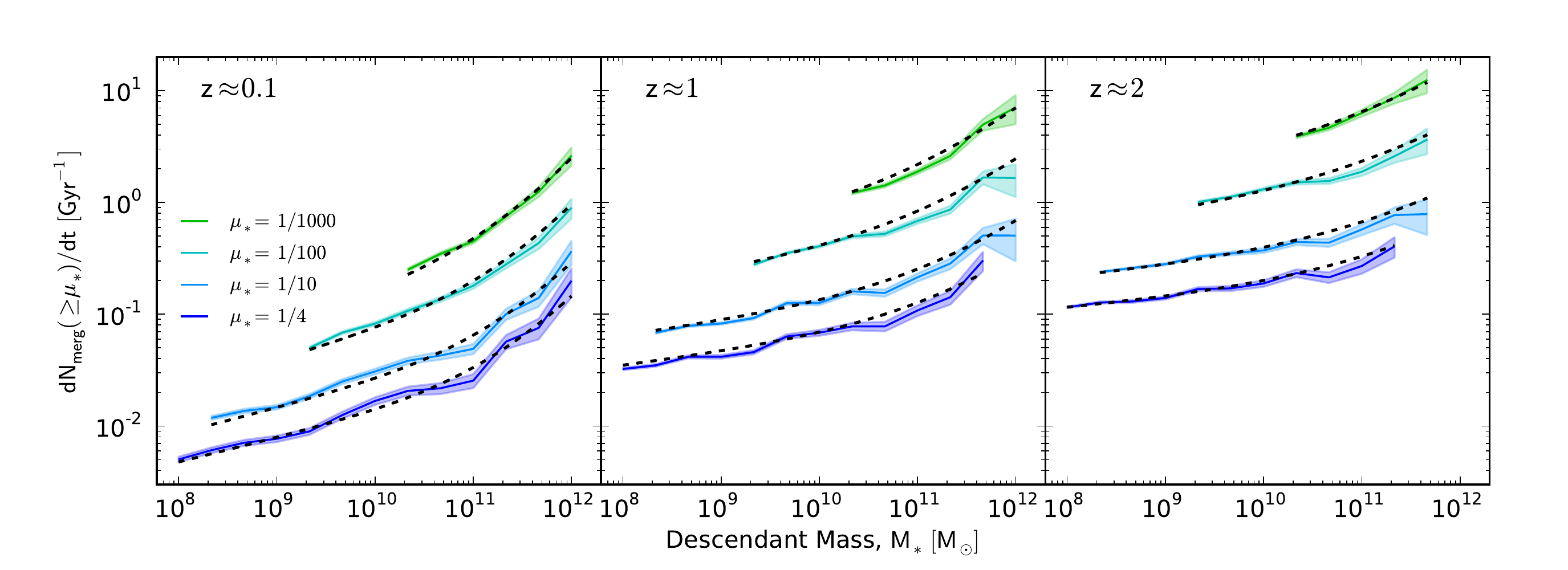}
			\includegraphics[width=\textwidth]{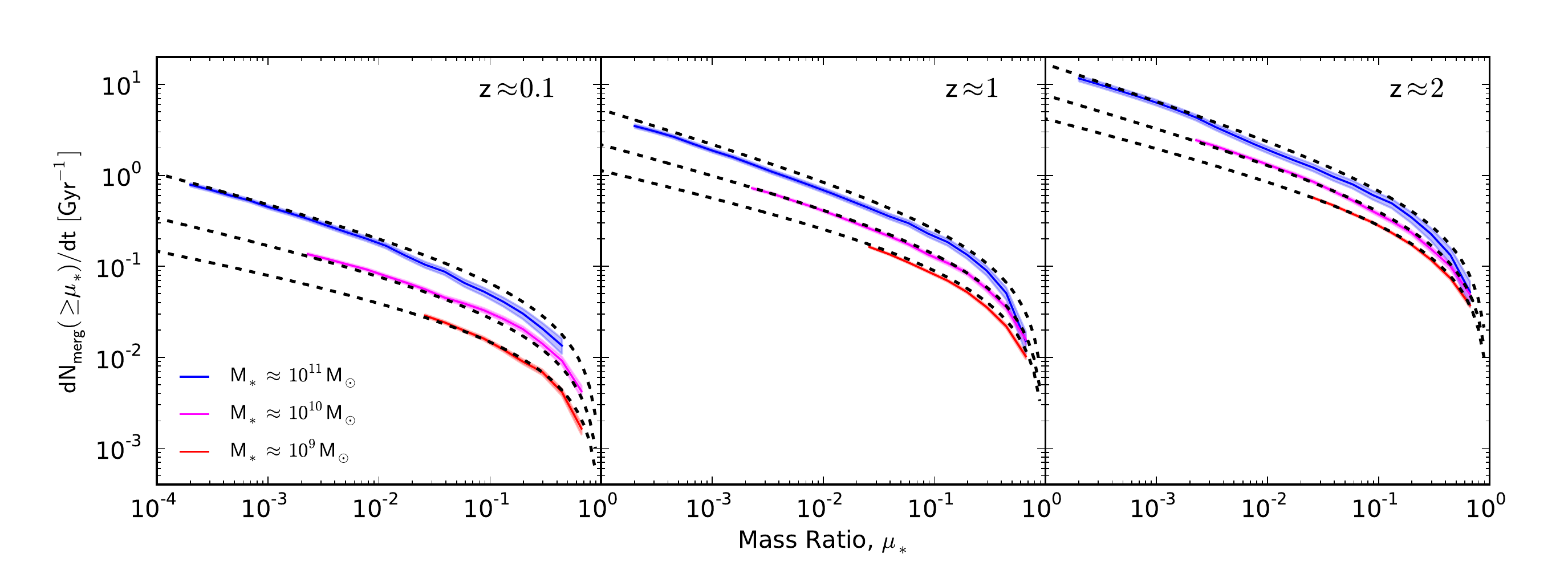}
		}
	}
\caption{\textit{Top:} The \textit{cumulative} (with respect to mass ratio) merger rate as a function of the descendant mass $M_{\ast}$, shown for different minimum mass ratios. \textit{Bottom:} The \textit{cumulative} merger rate as a function of the mass ratio $\mu_{\ast}$, shown for different descendant masses. The left, center, and right panels correspond to redshift bins centered around 0.1, 1, and 2, respectively. The shaded regions indicate the Poisson noise in the number of mergers in each bin. The black dashed line is the \textit{integral} of the fitting function from Table \ref{tab:fitting_formula}, integrated over the appropriate range of mass ratios $\mu_{\ast}$ (and therefore not a direct fit to the data shown in this figure).}
\label{fig:cumulative_merger_rate}
\end{figure*}

To address the time delay between $t_{\rm infall}$ and $t_{\rm max}$ more generally, Figure \ref{fig:mass_timescales} shows the elapsed time since $t_{\rm max}$ and since $t_{\rm infall}$ for all merging (left) and surviving (right) satellites at $z=0$. The bottom panels show the difference between the two times, which indicates that for the vast majority of satellites, $t_{\rm max}$ takes place a few Gyr after $t_{\rm infall}$. The difference between these two time-scales is more pronounced and shows a smaller scatter in the case of merging satellites, which is partly explained by the fact that the surviving satellite population (right) includes galaxies which have been more recently accreted onto the halo, shifting $\Delta t_{\rm infall}$ downward (i.e., infall takes place at a later time) relative to the merging satellite population. Furthermore, it is less likely that newly accreted satellites have undergone increased star formation due to interactions with other galaxies, which shifts $\Delta t_{\rm max}$ upward (i.e., the maximum stellar mass was reached earlier) relative to merging satellites which, by definition, have already undergone such interactions.

		All of this favors $t_{\rm max}$ over the other alternatives for the time when the merger mass ratio should be defined.\footnote{Perhaps another interesting alternative would consist of taking both progenitor masses at the time when the secondary enters the tidal radius of the interacting pair. However, such an alternative would be sensitive to the mass ratio between the primary and secondary progenitors, which, as we have seen, is largely influenced by details of the halo finding algorithm.} By this time most of the stellar mass of the galaxy has already been formed, but it is also before numerical and physical effects from the merger itself begin to dominate. In other words, by taking the progenitor masses at $t_{\rm max}$ we are minimizing the bias from two different effects that tend to underestimate the stellar mass of the secondary progenitor, although for different reasons. We conclude that taking the progenitor masses at $t_{\rm max}$ is a reasonable choice for calculating both the merger rate \textit{and} the stellar mass accretion rate, which will be the topic of upcoming work.

	\subsection{Results}\label{subsec:merger_rate_results}

		In this section we present the main features of the galaxy-galaxy merger rate as a function of descendant stellar mass $M_{\ast}$, progenitor stellar mass ratio $\mu_{\ast}$, and redshift $z$. We consider both the `differential' merger rate, which corresponds to mergers with mass ratios within a given interval, as well as the `cumulative' merger rate, which includes all mergers with mass ratios greater than a given minimum value.
		
		Figure \ref{fig:differential_merger_rate} shows the differential merger rate, given by Equation (\ref{eq:merger_rate}), as a function of descendant mass (top) and as a function of mass ratio (bottom). The panels from left to right correspond to redshift bins centered around $z = 0.1$, 1 and 2, respectively. The dashed black line corresponds to the fit from Table \ref{tab:fitting_formula}. We note that the merger rate has a relatively simple dependence on both $M_{\ast}$ and $\mu_{\ast}$. The dependence with respect to $\mu_{\ast}$ is well described by a power law, while the dependence on $M_{\ast}$ can be modeled with a double power law with a break around $M_{\ast} \approx 2 \times 10^{11} \, \Msun$. We note that the merger rate is always an increasing function of descendant galaxy mass: at low masses it grows as $\sim M_{\ast}^{0.2}$, which is very close to the mass dependence of the halo merger rate, $\sim M_{\rm halo}^{0.13-0.15}$ \citep{Genel2010, Fakhouri2010}, and it steepens at $M_{\ast} \gtrsim 2 \times 10^{11} \, \Msun$.
		
		Close inspection of the bottom panels of Figure \ref{fig:differential_merger_rate} reveals that the lines corresponding to different descendant masses are not exactly parallel to each other. This feature is modeled by a `mixed' term which includes both $M_{\ast}$ and $\mu_{\ast}$, and which is parametrized by $\gamma$ (see Table \ref{tab:fitting_formula}). This means that, unlike with the halo merger rate, the galaxy merger rate is not separable with respect to descendant mass and mass ratio. This feature of the merger rate implies that more massive galaxies have a slightly larger relative contribution from more minor mergers, compared to less massive galaxies.
		
		Figure \ref{fig:cumulative_merger_rate} is similar to Figure \ref{fig:differential_merger_rate}, except that the merger rate is now `cumulative' with respect to the mass ratio, i.e., it includes all mergers with mass ratios greater than a given $\mu_{\ast}$, and the fitting function has been integrated accordingly (strictly speaking, it is not a fit anymore). We note that the enhancement in the galaxy merger rate above $M_{\ast} \approx 2 \times 10^{11} \, \Msun$ becomes more noticeable after the merger rate has been integrated with respect to mass ratio. This feature is presumably a manifestation of the `turnover' in the $M_{\ast}-M_{\rm halo}$ relationship, as explained in \cite{Hopkins2010a}.

		Figure \ref{fig:cumulative_merger_rate} can be useful for making quick assessments of the number of mergers that galaxies of a certain stellar mass are expected to undergo during a given time interval. For example, the major merger rate at $z \approx 0.1$ (blue line, upper left panel) for Milky Way-like galaxies with $M_{\ast} \approx 6 \times 10^{10} \, \Msun$ \citep{McMillan2011} is slightly larger than 0.02 Gyr$^{-1}$, which means that roughly one in every 50 Milky Way-like galaxies has undergone a major merger during the last Gyr.
		
\begin{figure*}\centerline{\hbox{
\includegraphics[width=\columnwidth]{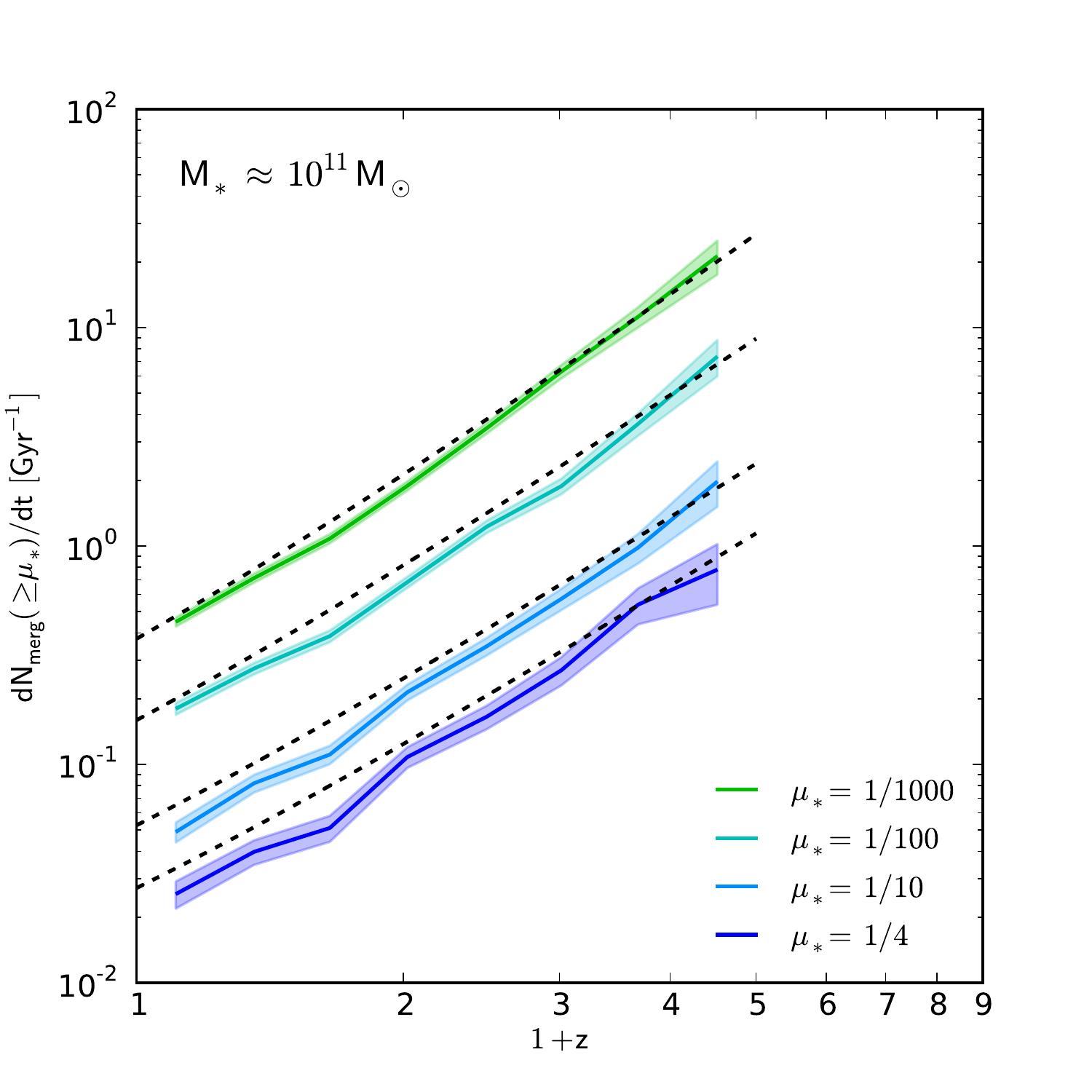}
\includegraphics[width=\columnwidth]{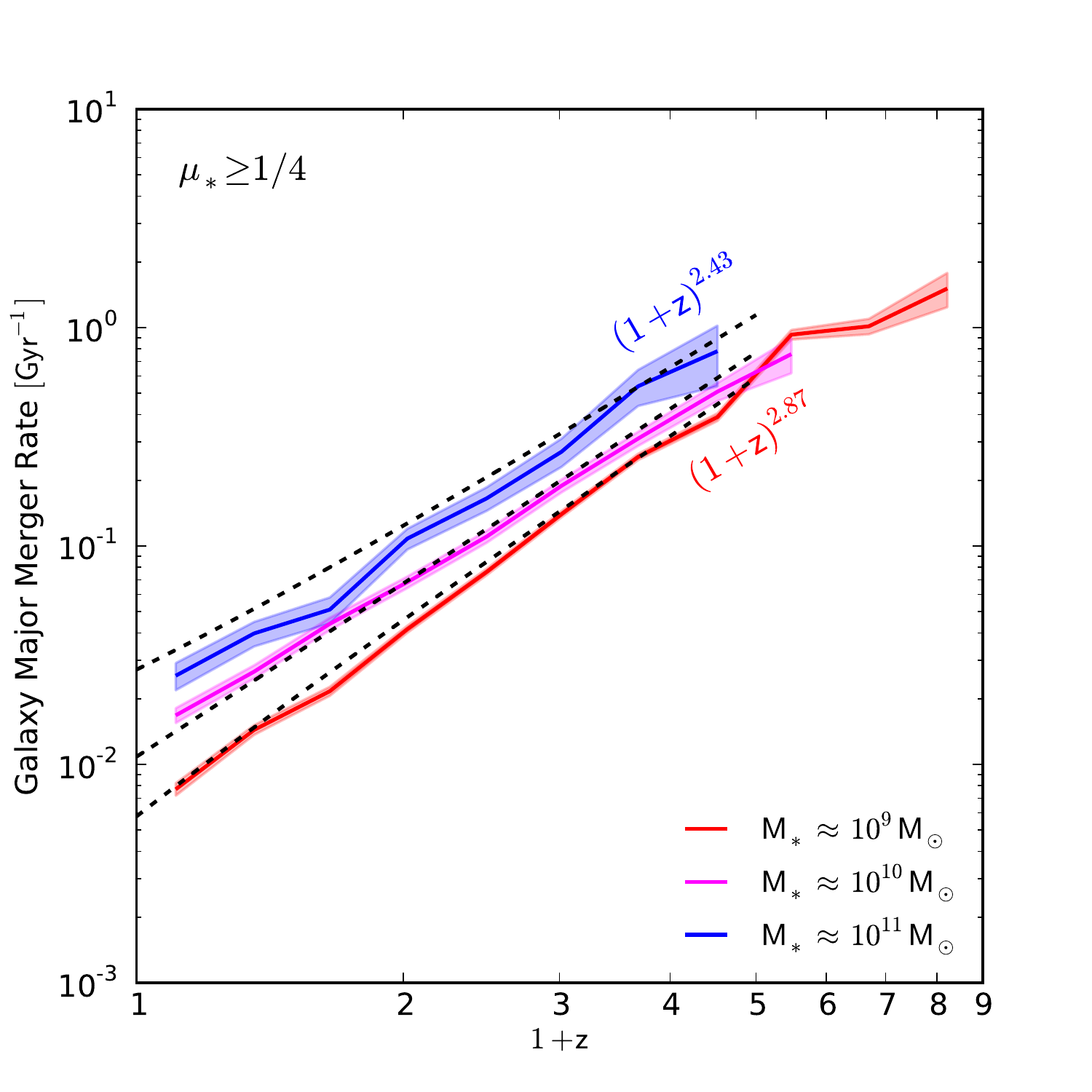}}}
\caption{\textit{Left:} The \textit{cumulative} (with respect to mass ratio) merger rate as a function of redshift for descendant masses $M_{\ast} \approx 10^{11} \, \Msun$, shown for a variety of minimum mass ratios. \textit{Right:} The \textit{cumulative} (with respect to mass ratio) merger rate as a function of redshift for mass ratios $\mu_{\ast} \geq 1/4$ (i.e., the major merger rate), shown for different descendant masses. The shaded regions indicate the Poisson noise in the number of mergers in each bin. The black dashed line represents the fitting function from Table \ref{tab:fitting_formula}, integrated over the appropriate mass ratio interval (it is therefore not a direct fit to the data shown in this figure).}
\label{fig:merger_rate_vs_redshift}
\end{figure*}

		Figure \ref{fig:merger_rate_vs_redshift} shows the redshift dependence of the cumulative (i.e., including all mergers with mass ratios larger than a given $\mu_{\ast}$) galaxy merger rate. The left panel shows the merger rate of galaxies with a fixed descendant mass $M_{\ast} \approx 10^{11} \, \Msun$ for different mass ratio thresholds, while the right panel shows the major ($\mu_{\ast} \geq 1/4$) merger rate for different descendant masses.

		The right panel from Figure \ref{fig:merger_rate_vs_redshift} demonstrates that the redshift dependence of the major merger rate becomes slightly weaker for more massive galaxies, as observed by \cite{Hopkins2010} using semi-empirical methods. We find that the major merger rate of $M_{\ast} \approx 10^9 \, \Msun$ galaxies has a redshift dependence proportional to $\sim (1+z)^{2.87}$, while the the major merger rate of $M_{\ast} \approx 10^{11} \, \Msun$ galaxies evolves as $\sim (1+z)^{2.43}$. On the other hand, the left panel from Figure \ref{fig:merger_rate_vs_redshift} shows that the slope of the merger rate with respect to redshift is practically independent of the mass ratio. In other words, the relative amount of major and minor mergers undergone by every galaxy (on average) is the same for all redshifts. In general, we find that the redshift dependence of the galaxy merger rate is very similar to the one of the halo merger rate, which evolves as $\sim (1+z)^{2.2-2.3}$ \citep{Genel2010, Fakhouri2010}.

	\subsection{Comparison to observations and semi-empirical models}\label{subsec:observations}

	\begin{figure*}\centerline{
	\hbox{
	\includegraphics[width=\columnwidth]{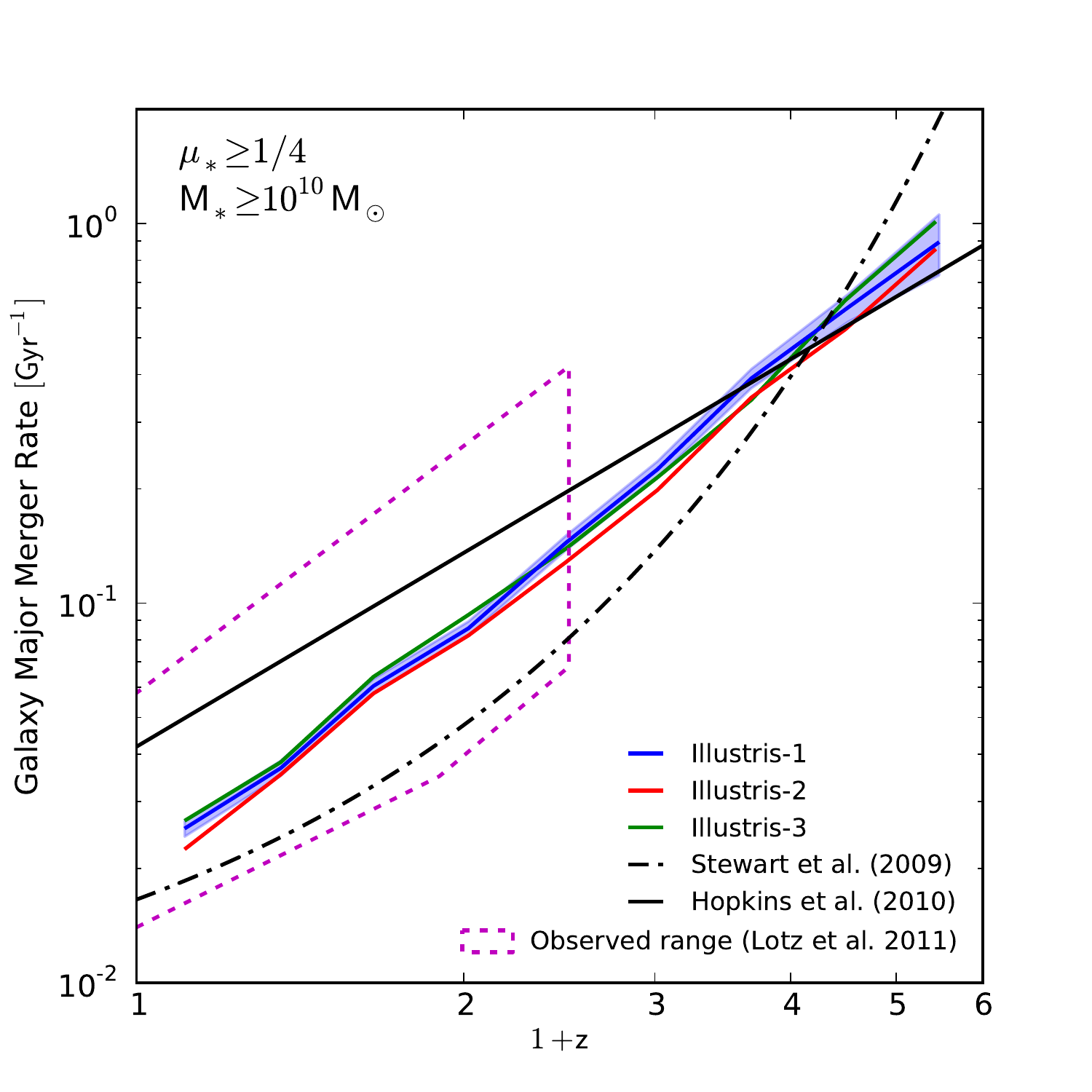}
	\includegraphics[width=\columnwidth]{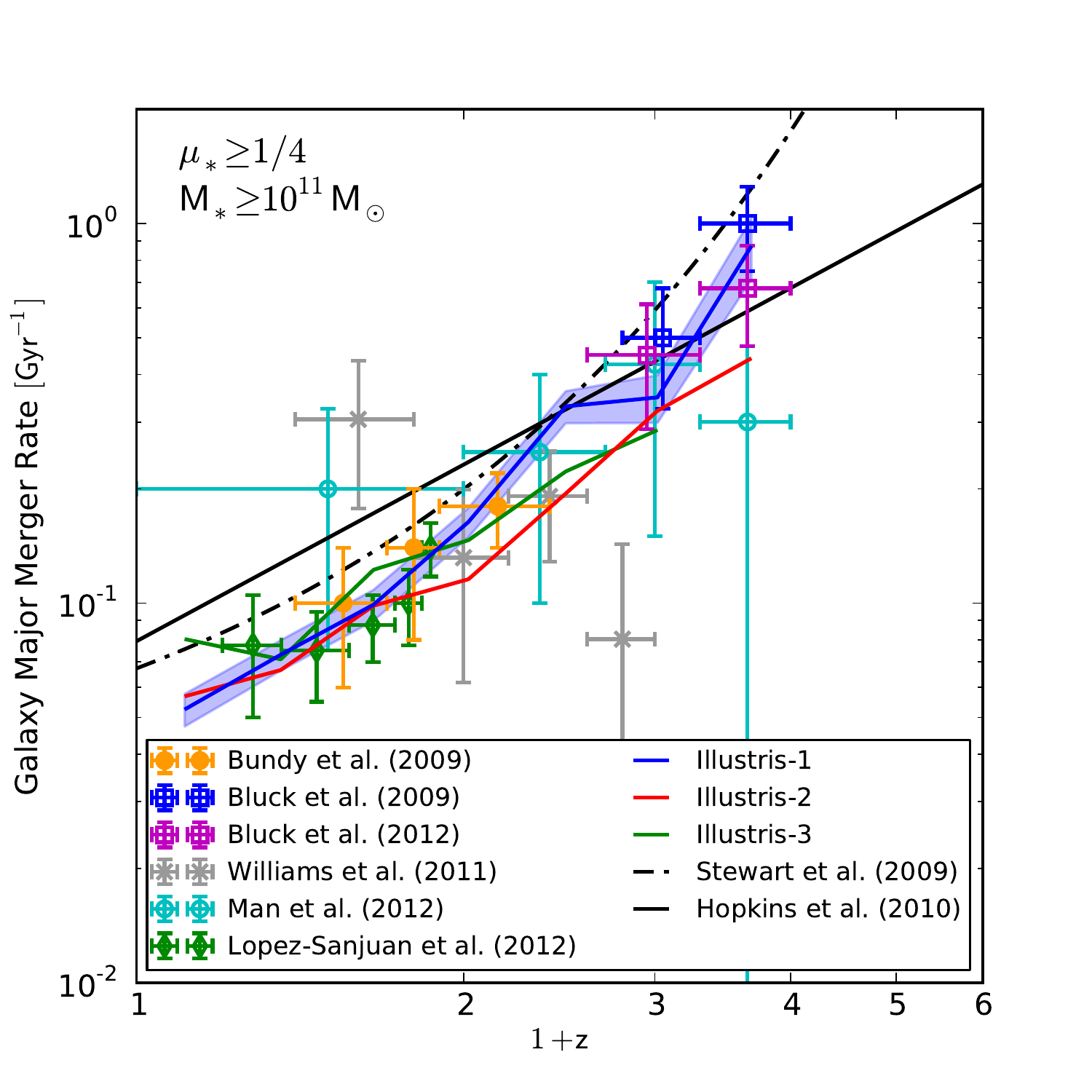}
	}}
	\caption{The galaxy major merger rate ($\mu_{\ast} \geq 1/4$) as a function of redshift, for descendant stellar masses greater than $10^{10} \, \Msun$ (left) and $10^{11} \, \Msun$ (right). The blue, red, and green lines correspond to the three resolution levels of Illustris. The shaded regions (Illustris-1 only) correspond to the Poisson noise from the number of mergers in each bin. Fitting functions from the semi-empirical models of \protect\cite{Stewart2009} and \protect\cite{Hopkins2010} are indicated with dot-dashed and solid black lines, respectively. The magenta dashed range on the left panel encapsulates the observational constraints for medium-sized galaxies ($M_{\ast} \gtrsim 10^{10} \, \Msun$), determined from observations of the merger fraction by \protect\cite{Kartaltepe2007}, \protect\cite{Lin2008}, \protect\cite{DeRavel2009}, and \protect\cite{Bundy2009}, in combination with cosmologically averaged merger time-scales from \protect\cite{Lotz2011}. The right panel includes different observational estimates of the merger rate for massive galaxies ($M_{\ast} \gtrsim 10^{11} \, \Msun$), shown as symbols with errorbars.}
	\label{fig:mr_vs_z_observations}
	\end{figure*}

	\begin{figure}
	\includegraphics[width=\columnwidth]{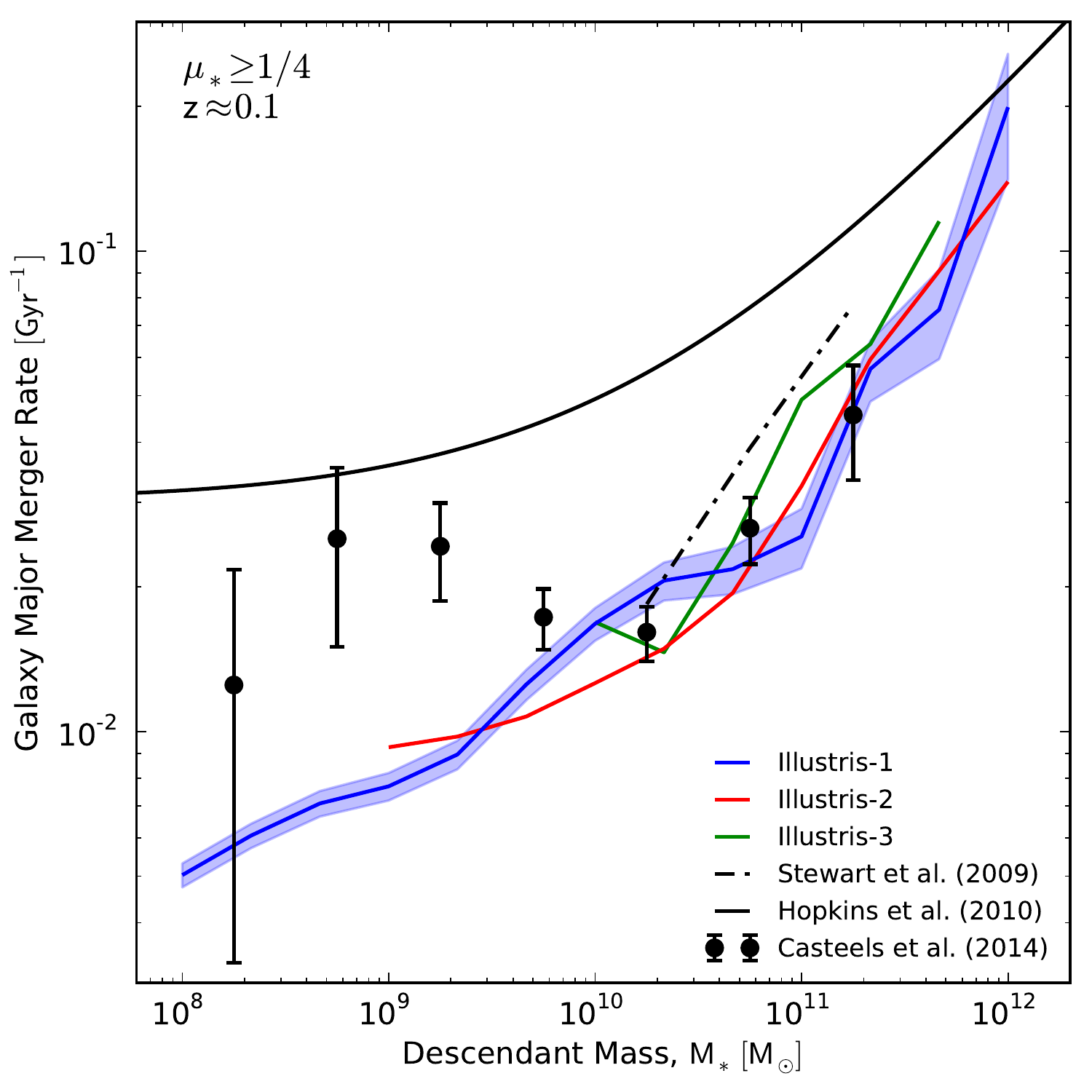}
	\caption{The galaxy major merger rate ($\mu_{\ast} \geq 1/4$) as a function of descendant stellar mass, for a redshift bin centered around $z = 0.1$. The blue, red, and green lines correspond to the three resolution levels of Illustris. The shaded regions (Illustris-1 only) correspond to the Poisson noise from the number of mergers in each bin. Fitting functions from the semi-empirical models of \protect\cite{Stewart2009} and \protect\cite{Hopkins2010}, evaluated at $z=0.1$, are indicated with dot-dashed and solid black lines, respectively. The model from \protect\cite{Hopkins2010} has been scaled so that it corresponds to a major merger definition of $\mu_{\ast} \geq 1/4$ instead of $\mu_{\ast} \geq 1/3$. The black circles with errorbars correspond to recent observations from \protect\cite{Casteels2014a}.}
	\label{fig:mr_vs_mass_observations}
	\end{figure}

		In this section we compare our main results with observational estimates of the galaxy merger rate, as well as with predictions from semi-empirical models. We do not include results from SAMs \citep[e.g.,][]{Guo2008} or hydrodynamic simulations \citep[e.g.,][]{Maller2006} because their differences with respect to observations and semi-empirical models have already been studied in \cite{Hopkins2010a}.

		Figure \ref{fig:mr_vs_z_observations} shows the major ($\mu_{\ast} \geq 1/4$) merger rate of medium-sized ($M_{\ast} \geq 10^{10} \, \Msun$, left) and massive ($M_{\ast} \geq 10^{11} \, \Msun$, right) galaxies as a function of redshift. The blue, red and green solid lines correspond to the different resolutions of Illustris, while the dot-dashed and solid black lines show predictions from the semi-empirical models of \cite{Stewart2009} and \cite{Hopkins2010}, respectively. These semi-empirical models disagree among themselves by factors of up to $\sim$2--3, and our results from Illustris generally lie within this uncertainty range.
		
		We point out that the galaxy merger rate in Figure \ref{fig:mr_vs_z_observations} is slightly different from the one in Figure \ref{fig:merger_rate_vs_redshift} because we now include all galaxies with stellar masses \textit{larger than} a given value, rather than \textit{around} a given value. This is done in order to have a more meaningful comparison with observations, which typically consider all galaxies with stellar masses (or luminosities) above a certain threshold. Additionally, the fitting functions from \cite{Stewart2009} and \cite{Hopkins2010} represent slightly different quantities. The one from \cite{Hopkins2010} describes the merger rate for all galaxies with masses \textit{larger than} a given value, while \cite{Stewart2009} provide three different versions of their fitting formula, with parameters corresponding to the mass ranges $10^{10} < M_{\ast}/\Msun < 10^{10.5}$ (shown in the left panel of Figure \ref{fig:mr_vs_z_observations}), $10^{10.5} < M_{\ast}/\Msun < 10^{11}$, and $M_{\ast} > 10^{11} \Msun$ (shown in the right panel of Figure \ref{fig:mr_vs_z_observations}). Since the galaxy merger rate in both of these models (as well as in the current work) is an increasing function of descendant mass, the fit by \cite{Stewart2009} on the left panel of Figure \ref{fig:mr_vs_z_observations} should be considered as a lower bound (although by less than 30 per cent, as a consequence of the weak mass dependence of the merger rate, and also because the number density of galaxies is dominated by less massive ones).

		The left panel of Figure \ref{fig:mr_vs_z_observations} also shows the range allowed by observations according to theoretical work by \cite{Lotz2011}, where observational estimates of the major merger fraction from \cite{Kartaltepe2007}, \cite{Lin2008}, \cite{DeRavel2009}, and \cite{Bundy2009} are converted into merger rates by means of `cosmologically averaged' observability time-scales, which are determined from hydrodynamic merger simulations in combination with a galaxy formation model \citep{Somerville2008}. The corresponding galaxy merger rates predicted by Illustris are in good agreement with the predictions from \cite{Lotz2011}, as well as with the semi-empirical models of \cite{Stewart2009} and \cite{Hopkins2010a}, which are all allowed by the observational constraints.


		The right panel of Figure \ref{fig:mr_vs_z_observations}, which corresponds to more massive galaxies ($M_{\ast} \geq 10^{11} \, \Msun$), includes observational estimates of the merger rate based on merger fraction measurements by \cite{Bundy2009}, \cite{Bluck2009, Bluck2012}, \cite{Williams2011}, \cite{Man2012}, and \cite{Lopez-Sanjuan2012}, which are shown as symbols with errorbars. In all cases we adopt the merger time-scales suggested by the authors, which are typically between 0.4 and 0.5 Gyr, except for the pair fraction observations of \cite{Williams2011} and \cite{Lopez-Sanjuan2012}, where we adopt a time-scale of 0.4 Gyr instead of the significantly larger suggested time-scales. Other observations of the merger fraction for massive galaxies which are not shown have been carried out by \cite{deRavel2011}, \cite{Newman2012}, \cite{Xu2012}, \cite{Ferreras2014}, and \cite{Lackner2014}.
		
		In the case of massive galaxies, different authors find qualitatively different trends in the redshift evolution of the merger fraction: a decreasing redshift dependence \citep{Williams2011, Ferreras2014}, a nearly constant or mildly increasing redshift dependence \citep{deRavel2011, Man2012, Newman2012, Xu2012}, or a strongly increasing redshift dependence \citep{Bundy2009, Bluck2009, Bluck2012, Lopez-Sanjuan2012, Lackner2014}. Recently, \cite{Man2014} attempted to resolve some of these differences by pointing out that studies in which major mergers are selected by flux ratio instead of stellar mass ratio tend to include very bright galaxies which nevertheless have very small masses, and should therefore not be counted as major mergers (using a stellar mass ratio definition). Therefore, \cite{Man2014} also support a decreasing redshift dependence, assuming that major mergers are selected by their stellar mass ratio. Yet, the Illustris Simulation (as well as semi-empirical models) predict a strongly increasing redshift dependence, despite the fact that major mergers are also selected by stellar mass ratio.

		The reason for this discrepancy is unclear at this stage. On the one hand, until observations converge to an agreed result better than a factor of $\sim$2, they will not be able to place significant constraints on modern theoretical models. On the other hand, considering that the halo-halo merger rate also exhibits a strong, positive correlation with redshift, we cannot envision any physical mechanism for which such trend should reverse in the case of galaxy mergers.
		
		Figure \ref{fig:mr_vs_mass_observations} shows the major ($\mu_{\ast} \geq 1/4$) merger rate of galaxies as a function of descendant stellar mass. As before, the three resolutions of Illustris are indicated with blue, red and green solid lines, and predictions from the semi-empirical models of \cite{Stewart2009} and \cite{Hopkins2010} are shown with dot-dashed and solid black lines, respectively. The black circles with errorbars correspond to recent observational work on the mass-dependent merger rate by \cite{Casteels2014a}, based on observations of the fraction of highly asymmetric galaxies in the local Universe (z $\lesssim$ 0.2), which are converted into merger rates by using the mass-dependent merger time-scales from \cite{Conselice2006}. The galaxy merger rate in Illustris is in good agreement with the observations by \cite{Casteels2014a} for galaxies with stellar masses $M_{\ast} \gtrsim 10^{10} \, \Msun$, although there is some disagreement below $\sim 10^{10} \, \Msun$. We point out that the time-scales used by \cite{Casteels2014a} require gas fraction measurements, which are only available for $M_{\ast} > 10^{10} \, \Msun$. Therefore, an extrapolation has been used for $M_{\ast} < 10^{10} \, \Msun$, which can introduce significant uncertainties into the corresponding observability time-scales.

		The predictions from \cite{Hopkins2010} appear to be larger than the ones from Illustris by a factor of $\sim$2--5. Part of this difference is explained by the fact that the model from \cite{Hopkins2010} describes the merger rate for all galaxies with masses \textit{larger than} a given value, while the other estimates in Figure \ref{fig:mr_vs_mass_observations} correspond to galaxies with stellar masses \textit{around} a given value. According to calculations with Illustris, this can account for a factor of $\sim$2 at the low-mass end, $M_{\ast} \lesssim 10^9 \, \Msun$, but the effects become less significant at higher masses. The remaining differences are possibly related to the merger time-scales involved, which are included self-consistently in Illustris (see Section \ref{sec:discussion}).
		
		Finally, we point out that different observational estimates for the mass dependence of the galaxy merger rate have also not converged yet, with some studies supporting an increasing mass dependence and others suggesting the opposite \citep[see][for a discussion]{Casteels2014a}. In fact, the observations by \cite{Casteels2014a} are consistent with both an increasing and a decreasing mass dependence, depending on the stellar mass range considered. The Illustris Simulation, on the other hand, always predicts an increasing mass dependence, which becomes steeper for larger galaxy masses.

  \subsection{A fitting formula}\label{subsec:fitting_formula}

	In Table \ref{tab:fitting_formula} we provide a fitting formula for the galaxy-galaxy merger rate, along with the corresponding best-fitting parameters. For the sake of readability, we have dropped the asterisk subscript from the symbols $M_{\ast}$ and $\mu_{\ast}$. All masses and mass ratios in this section correspond to stellar masses.

		We find that the galaxy-galaxy merger rate has a relatively simple dependence on the descendant mass $M$, the progenitor mass ratio $\mu$, and the redshift $z$. The expression from Table \ref{tab:fitting_formula} is qualitatively similar to the fitting function for DM halo merger rates presented in \cite{Fakhouri2008}, which is essentially a power law in $M$, $\mu$ and $(1+z)$.

		The main difference between the mathematical forms of the halo-halo and galaxy-galaxy merger rates is that the mass dependence steepens significantly at the high-mass end in the case of galaxies, such that it is better described by a double power law with a break around $2 \times 10^{11} \, \Msun$. Furthermore, the exponents $\alpha$ and $\delta$ of the double power law exhibit some redshift dependence, which we parametrize as $\alpha(z) = \alpha_0 (1+z)^{\alpha_1}$ and $\delta(z) = \delta_0 (1+z)^{\delta_1}$. Both $\alpha_1$ and $\delta_1$ are negative, which means that the mass dependence of the merger rate weakens with increasing redshift. This also means, as mentioned earlier, that the redshift dependence is stronger for lower-mass galaxies. Additionally, the expression from Table \ref{tab:fitting_formula} contains a `mixed' term, parametrized by $\gamma$, that depends on both the stellar mass $M$ and the mass ratio $\mu$. This shows that the galaxy merger rate is not fully separable with respect to these variables, even for a fixed redshift.

		The fits were carried out in log-space by minimizing a chi-squared merit function with a Markov Chain Monte Carlo algorithm \citep[][http://dan.iel.fm/emcee/current/]{Foreman-Mackey2013}, considering all mergers which satisfy $M \geq 10^8 \, \Msun$, $\mu \geq 1/1000$, and $z \leq 4$. The data points were obtained by creating bins in $M$, $\mu$ and $(1+z)$ with widths corresponding to factors of 2, 1.2, and 1.1, respectively, and calculating the merger rate, along with the associated uncertainties, as explained in Section \ref{subsec:merger_rate_definitions}. In some cases the bins were rearranged so that there were at least 5 mergers per bin.

		In all cases, the MCMC algorithm produced approximately gaussian marginal distributions for each parameter. Therefore, we define the best-fitting value of each parameter as the mode of its marginal distribution, and the associated uncertainty as half the interval between the 16th and 84th percentiles, which corresponds to approximately $1\sigma$. The resulting best-fitting parameters yield a reduced chi-squared statistic with a value of 1.16, which indicates that the model from Table \ref{tab:fitting_formula} is a reasonably good fit to the data, without overfitting it.

\section{Discussion and Conclusions} \label{sec:discussion}

We have developed a theoretical framework for constructing and analyzing merger trees of galaxies and DM halos, which we apply to the Illustris Simulation \citep{Vogelsberger2014a, Vogelsberger2014, Genel2014a} to make theoretical predictions for the merger rates of galaxies and DM halos.

We find that the overall properties of DM halo merger trees and rates, which have been computed using the \textit{splitting} method \citep{Genel2009, Genel2010, Fakhouri2009a, Fakhouri2010a}, are robust to baryonic effects and are also in very good agreement with previous theoretical work by \cite{Genel2010}, who provided a fitting formula with parameters tuned to the Millennium and Millennium II simulations \citep{Springel2005b, Boylan-Kolchin2009}. This agreement shows that the volume covered by the Illustris simulation can be considered to be `representative' of the large-scale density field of the Universe.

The most novel aspect of this work pertains to the galaxy-galaxy merger rate, which we determine with unprecedented precision using a cosmological hydrodynamic simulation. We construct galaxy merger trees using an algorithm that has been shown to be reliable under a wide variety of circumstances \citep{Srisawat2013a, Avila2014, Lee2014}. In particular, our merger trees are designed to track the innermost regions of subhalos and galaxies, feature a robust definition of the first progenitor (i.e., the main progenitor), and avoid flyby events to some extent by allowing some objects to `skip' a snapshot when finding a descendant.

When calculating galaxy merger rates, we argue that the most meaningful definition of the merger mass ratio consists in taking the two progenitor masses at the moment when the secondary progenitor reaches its maximum stellar mass. This happens, on average, a few Gyr \textit{after} the secondary progenitor infalls into the same FoF group as the main progenitor. Additionally, we only consider mergers which have a well-defined \textit{infall} moment, as explained in Section \ref{subsec:merger_rate_definitions}. These definitions result in merger rates that are very well converged with resolution, as we show in Figures \ref{fig:mr_vs_z_observations} and \ref{fig:mr_vs_mass_observations}.

We find that the galaxy merger rate has a relatively simple dependence on descendant stellar mass, progenitor stellar mass ratio, and redshift, which is described by the fitting function given in Table \ref{tab:fitting_formula}. Essentially, this fit consists of a double power law with respect to stellar mass with a break around $\sim 2\times 10^{11} \, \Msun$, and single power laws for the mass ratio and redshift dependences. Some of the power law exponents change with redshift, which results in a mass dependence that weakens with increasing redshift, or, equivalently, a redshift dependence that weakens with increasing mass. There is also a clear correlation between descendant mass and progenitor mass ratio, even at a fixed redshift, which implies that the galaxy-galaxy merger rate is not separable with respect to these variables, in contrast with the mathematical form of the halo-halo merger rate \citep[e.g.][]{Fakhouri2008}.

The strong, positive correlation with redshift found in this work is in disagreement with some observations of the major merger fraction for massive galaxies ($M_{\ast} \gtrsim 10^{11} \, \Msun$), which find a nearly constant or decreasing evolution with redshift \citep{Williams2011, Newman2012, Man2014}. On the other hand, our results are in reasonable agreement with observations that suggest an increasing redshift evolution \citep{Bundy2009, Bluck2009, Bluck2012, Man2012}.

For medium-sized galaxies ($M_{\ast} \gtrsim 10^{10} \, \Msun$), the galaxy merger rate in Illustris is consistent with the general observational picture \citep[e.g.,][]{Lotz2011}. However, observational estimates of the merger rate must converge to a factor better than $\sim$2 in order to distinguish predictions based on semi-empirical models \citep{Stewart2009, Hopkins2010} -- which disagree among themselves by factors of up to $\sim$2--3 -- from those of Illustris, which typically lie inside this uncertainty range.

Observational work on the mass dependence of the merger rate has also not converged. We find good agreement with \cite{Casteels2014a} for galaxies with stellar masses above $\sim 10^{10} \, \Msun$, but find tension towards lower masses. This is possibly due to uncertainties in the observability time-scales assumed by \cite{Casteels2014a}, which require extrapolating the gas fraction for galaxies with stellar masses below $\sim 10^{10} \, \Msun$, where observational data is unavailable.

As already mentioned, the galaxy merger rate in Illustris is in good qualitative agreement with predictions from semi-empirical models \citep{Stewart2009, Hopkins2010a}. However, it is worth noting that such models are designed to give reasonable agreement with observations by construction, without attempting to model galaxy formation from first principles. Because of this, they cannot be used to study the dependence of the galaxy merger rate and related quantities with respect to variations in physical models of galaxy formation. Additionally, semi-empirical models are generally not applicable in situations where observational data is scarce, such as for making predictions for the merger rate in the very minor merger regime ($\mu_{\ast} \lesssim 1/10$), at high redshifts ($z \gtrsim 3$), or when measurements of gas fractions are required.

State-of-the-art cosmological hydrodynamic simulations are better suited for such tasks. Furthermore, they have the advantage of handling merger time-scales self-consistently, which makes them ideal for measuring the galaxy-galaxy merger rate. For example, merger time-scales can be complicated by interactions with a third external object, which appears to be a fairly common occurrence \citep{Moreno2013}. Additionally, the final stages of a major merger are dominated by loss of angular momentum due to baryonic resonances and tidal torques \citep[see][for a review]{Hayward2014}, which are difficult -- or impossible -- to describe using simple prescriptions for merger time-scales.

Previous attempts to measure the galaxy-galaxy merger rate using hydrodynamic simulations have yielded results which are significantly different from the ones presented in this work. In particular, the merger rate found by \cite{Maller2006} shows a much stronger dependence on descendant mass and redshift, which results in relatively poor agreement with observations and semi-empirical methods, as discussed in \cite{Hopkins2010a}. More recently, \cite{Kaviraj2014} calculated the galaxy merger rate in the Horizon-AGN cosmological hydrodynamical simulation \citep{Dubois2014} and found a nearly constant evolution with redshift, in disagreement with the results found by \cite{Maller2006}, as well as with the ones from Illustris. These differences can be driven by various factors, including differences in star formation physics and AGN feedback (or lack thereof), details about the substructure finding algorithm and merger tree construction method, or different definitions when calculating the merger rate, most notably the progenitor mass ratio. In general, estimating the galaxy-galaxy merger rate using \textit{a priori} models of galaxy formation is a non-trivial task.

The results presented in this paper are also in stark contrast with those found by \cite{Guo2008}, who applied the SAM proposed by \cite{DeLucia2007} to the Millennium Simulation \citep{Springel2005b} and found that the galaxy-galaxy merger rate has a strong dependence on stellar mass, but a weak one on redshift. In contrast, we find that it has a relatively weak dependence on stellar mass, but a strong one on redshift, which makes the galaxy-galaxy merger rate \textit{qualitatively} similar to the halo-halo merger rate (except for the `knee' in the mass dependence and the other features mentioned in Section \ref{subsec:fitting_formula}). Interestingly, \cite{Guo2008} also present an estimate of the halo-halo merger rate which is consistent with other theoretical calculations, including the one in this work. This implies that \cite{Guo2008} find large \textit{qualitative} differences between halo-halo and galaxy-galaxy merger rates, in disagreement with this work and with semi-empirical models. Some of these differences appear to be caused by satellite-specific prescriptions in the SAM of \cite{DeLucia2007}, in particular that galaxies cannot accrete gas after they have become satellites, and therefore cease to form stars once their supply of cold gas has been depleted.

The generally good agreement between the galaxy merger rate in Illustris and the one implied by observations comes with an important caveat: in order to convert a merger \textit{fraction} into a merger \textit{rate}, an observability time-scale has to be applied. This time-scale can shift the merger rate `vertically' to larger or smaller values, introducing some arbitrariness in its normalization. Up to now, the most accurate observability time-scales have been determined from hydrodynamic merger simulations \citep[e.g.,][]{Lotz2011}, averaged in a `cosmological context' by making several assumptions. Most importantly, a model of galaxy formation must be adopted in order to assign weights to the distribution of merger parameters at each redshift. Additionally, such merger simulations are usually considered to be in isolation, but, as mentioned above, \cite{Moreno2013} show that interactions between pairs of galaxies are often complicated by a third external object. These simplifying assumptions have a non-negligible effect on the merging time-scales of close pairs of galaxies, which affects estimates of the merger rate proportionately.

A more direct comparison with observations would consist of measuring the close pair fraction directly from the simulation, which could then be compared to observations without having to make assumptions about the merger time-scales involved. Unfortunately, this approach is complicated by the halo finder in situations where two large galaxies are found at very small separations ($\lesssim 20h^{-1}$ kpc). Ultimately, the best approach may consist in creating synthetic images of galaxy surveys using the Illustris simulation \citep[][Snyder et al., in preparation]{Torrey2014a} and then applying the same source identification algorithms that are used with observational images. These topics will be explored in upcoming work.

Whereas it is reassuring that the \textit{normalization} of the galaxy merger rate obtained in this work appears to agree well with observations, perhaps a more convincing indication of agreement is that the \textit{slope} of the merger rate as a function of redshift follows the same trend as the range allowed by observational constraints for medium-sized galaxies ($M_{\ast} \gtrsim 10^{10} \, \Msun$), which is proportional to $\sim (1+z)^{2.2-2.5}$. Although we cannot make a similar statement for more massive galaxies ($M_{\ast} \gtrsim 10^{11} \, \Msun$) due to the \textit{qualitative} disagreement between different observations of the merger fraction (Figure \ref{fig:mr_vs_z_observations}, right panel), the agreement with at least some of the sets of observations is encouraging. Additionally, the slope of the galaxy merger rate with respect to descendant mass is in good agreement with recent observations by \cite{Casteels2014a} for galaxies with stellar masses above $\sim 10^{10}\, \Msun$, where the observability time-scales used are more reliable. The body of these results shows that the Illustris Simulation can be used to make realistic predictions about galaxy merger rates and related quantities. Further work on merger time-scales and mock galaxy surveys will lead to even more detailed comparisons between theoretical models of galaxy formation and observations of interacting and morphologically disturbed galaxies.

\section*{Acknowledgments}

We are grateful to Asa Bluck, Gurtina Besla, Sarah Wellons, and Joshua Suresh for useful comments and discussions. We also thank Laura Blecha for carefully reading the manuscript. VS acknowledges support by the European Research Council through ERC-StG grant EXAGAL-308037. LH acknowledges support from NASA grant NNX12AC67G and NSF grant AST-1312095.

\bibliographystyle{mn2eFixed}

\bibliography{paper}

\begin{thebibliography}{69}
\expandafter\ifx\csname natexlab\endcsname\relax\def\natexlab#1{#1}\fi

\bibitem[{Avila {et~al}\mbox{.}(2014)Avila, Knebe, Pearce, Schneider, Srisawat,
  Thomas, Behroozi, Elahi, Han, Mao, Onions, Rodriguez-Gomez, \&
  Tweed}]{Avila2014}
Avila S. {et~al.}, 2014, Monthly Notices of the Royal Astronomical Society,
  441, 3488

\bibitem[{Bluck {et~al}\mbox{.}(2009)Bluck, Conselice, Bouwens, Daddi,
  Dickinson, Papovich, \& Yan}]{Bluck2009}
Bluck A. F.~L., Conselice C.~J., Bouwens R.~J., Daddi E., Dickinson M.,
  Papovich C., Yan H., 2009, Monthly Notices of the Royal Astronomical Society:
  Letters, 394, L51

\bibitem[{Bluck {et~al}\mbox{.}(2012)Bluck, Conselice, Buitrago,
  Gr\"{u}tzbauch, Hoyos, Mortlock, \& Bauer}]{Bluck2012}
Bluck A. F.~L., Conselice C.~J., Buitrago F., Gr\"{u}tzbauch R., Hoyos C.,
  Mortlock A., Bauer A.~E., 2012, The Astrophysical Journal, 747, 34

\bibitem[{Bond {et~al}\mbox{.}(1991)Bond, Cole, Efstathiou, \&
  Kaiser}]{Bond1991}
Bond J.~R., Cole S., Efstathiou G., Kaiser N., 1991, The Astrophysical
  JournalThe Astrophysical Journal, 379, 440

\bibitem[{Boylan-Kolchin {et~al}\mbox{.}(2009)Boylan-Kolchin, Springel, White,
  Jenkins, \& Lemson}]{Boylan-Kolchin2009}
Boylan-Kolchin M., Springel V., White S. D.~M., Jenkins A., Lemson G., 2009,
  Monthly Notices of the Royal Astronomical Society, 398, 1150

\bibitem[{Bundy {et~al}\mbox{.}(2009)Bundy, Fukugita, Ellis, Targett, Belli, \&
  Kodama}]{Bundy2009}
Bundy K., Fukugita M., Ellis R.~S., Targett T.~a., Belli S., Kodama T., 2009,
  The Astrophysical Journal, 697, 1369

\bibitem[{Casteels {et~al}\mbox{.}(2014)Casteels, Conselice, Bamford,
  Salvador-Sole, Norberg, Agius, Baldry, Brough, Brown, Drinkwater, Driver,
  Graham, Bland-Hawthorn, Hopkins, Kelvin, Lopez-Sanchez, Loveday, Robotham, \&
  Vazquez-Mata}]{Casteels2014a}
Casteels K. R.~V. {et~al.}, 2014, Monthly Notices of the Royal Astronomical
  Society, 445, 1157

\bibitem[{Conselice(2006)}]{Conselice2006}
Conselice C.~J., 2006, The Astrophysical Journal, 638, 686

\bibitem[{Davis {et~al}\mbox{.}(1985)Davis, Efstathiou, Frenk, \&
  White}]{Davis1985}
Davis M., Efstathiou G., Frenk C.~S., White S. D.~M., 1985, The Astrophysical
  Journal, 292, 371

\bibitem[{{De Lucia} \& Blaizot(2007)}]{DeLucia2007}
{De Lucia} G., Blaizot J., 2007, Monthly Notices of the Royal Astronomical
  Society, 375, 2

\bibitem[{de~Ravel {et~al}\mbox{.}(2011)de~Ravel, Kampczyk, F\`{e}vre, Lilly,
  Tasca, Tresse, Lopez-Sanjuan, Bolzonella, Kovac, Abbas, Bardelli, Bongiorno,
  Caputi, Contini, Coppa, Cucciati, de~la Torre, Dunlop, Franzetti, Garilli,
  Iovino, Kneib, Koekemoer, Knobel, Lamareille, Borgne, Brun, Leauthaud, Maier,
  Mainieri, Mignoli, Pello, Peng, Montero, Ricciardelli, Scodeggio, Silverman,
  Tanaka, Vergani, Zamorani, Zucca, Bottini, Cappi, Carollo, Cassata, Cimatti,
  Fumana, Guzzo, Maccagni, Marinoni, McCracken, Memeo, Meneux, Oesch, Porciani,
  Pozzetti, Renzini, Scaramella, \& Scarlata}]{deRavel2011}
de~Ravel L. {et~al.}, 2011, preprint (arXiv:1104.5470), 17

\bibitem[{de~Ravel {et~al}\mbox{.}(2009)de~Ravel, {Le F\`{e}vre}, Tresse,
  Bottini, Garilli, {Le Brun}, Maccagni, Scaramella, Scodeggio, Vettolani,
  Zanichelli, Adami, Arnouts, Bardelli, Bolzonella, Cappi, Charlot, Ciliegi,
  Contini, Foucaud, Franzetti, Gavignaud, Guzzo, Ilbert, Iovino, Lamareille,
  McCracken, Marano, Marinoni, Mazure, Meneux, Merighi, Paltani, Pell\`{o},
  Pollo, Pozzetti, Radovich, Vergani, Zamorani, Zucca, Bondi, Bongiorno,
  Brinchmann, Cucciati, de~la Torre, Gregorini, Memeo, Perez-Montero, Mellier,
  Merluzzi, \& Temporin}]{DeRavel2009}
de~Ravel L. {et~al.}, 2009, Astronomy and Astrophysics, 498, 379

\bibitem[{Dolag {et~al}\mbox{.}(2009)Dolag, Borgani, Murante, \&
  Springel}]{Dolag2009a}
Dolag K., Borgani S., Murante G., Springel V., 2009, Monthly Notices of the
  Royal Astronomical Society, 399, 497

\bibitem[{Dubois {et~al}\mbox{.}(2014)Dubois, Pichon, Welker, {Le Borgne},
  Devriendt, Laigle, Codis, Pogosyan, Arnouts, Benabed, Bertin, Blaizot,
  Bouchet, Cardoso, Colombi, de~Lapparent, Desjacques, Gavazzi, Kassin, Kimm,
  McCracken, Milliard, Peirani, Prunet, Rouberol, Silk, Slyz, Sousbie,
  Teyssier, Tresse, Treyer, Vibert, \& Volonteri}]{Dubois2014}
Dubois Y. {et~al.}, 2014, Monthly Notices of the Royal Astronomical Society,
  444, 1453

\bibitem[{Fakhouri \& Ma(2008)}]{Fakhouri2008}
Fakhouri O., Ma C.-P., 2008, Monthly Notices of the Royal Astronomical Society,
  386, 577

\bibitem[{Fakhouri \& Ma(2009)}]{Fakhouri2009a}
Fakhouri O., Ma C.-P., 2009, Monthly Notices of the Royal Astronomical Society,
  394, 1825

\bibitem[{Fakhouri \& Ma(2010)}]{Fakhouri2010a}
Fakhouri O., Ma C.-P., 2010, Monthly Notices of the Royal Astronomical Society,
  401, 2245

\bibitem[{Fakhouri, Ma \& Boylan-Kolchin(2010)Fakhouri, Ma, \&
  Boylan-Kolchin}]{Fakhouri2010}
Fakhouri O., Ma C.-P., Boylan-Kolchin M., 2010, Monthly Notices of the Royal
  Astronomical Society, 406, 2267

\bibitem[{Ferreras {et~al}\mbox{.}(2014)Ferreras, Trujillo,
  M\'{a}rmol-Queralt\'{o}, P\'{e}rez-Gonz\'{a}lez, Cava, Barro, Cenarro,
  Hern\'{a}n-Caballero, Cardiel, Rodr\'{\i}guez-Zaur\'{\i}n, \&
  Cebri\'{a}n}]{Ferreras2014}
Ferreras I. {et~al.}, 2014, Monthly Notices of the Royal Astronomical Society,
  444, 906

\bibitem[{Foreman-Mackey {et~al}\mbox{.}(2013)Foreman-Mackey, Hogg, Lang, \&
  Goodman}]{Foreman-Mackey2013}
Foreman-Mackey D., Hogg D.~W., Lang D., Goodman J., 2013, Publications of the
  Astronomical Society of the Pacific, 125, 306

\bibitem[{Genel {et~al}\mbox{.}(2010)Genel, Bouch\'{e}, Naab, Sternberg, \&
  Genzel}]{Genel2010}
Genel S., Bouch\'{e} N., Naab T., Sternberg A., Genzel R., 2010, The
  Astrophysical Journal, 719, 229

\bibitem[{Genel {et~al}\mbox{.}(2009)Genel, Genzel, Bouch\'{e}, Naab, \&
  Sternberg}]{Genel2009}
Genel S., Genzel R., Bouch\'{e} N., Naab T., Sternberg A., 2009, The
  Astrophysical Journal, 701, 2002

\bibitem[{Genel {et~al}\mbox{.}(2013)Genel, Vogelsberger, Nelson, Sijacki,
  Springel, \& Hernquist}]{Genel2013}
Genel S., Vogelsberger M., Nelson D., Sijacki D., Springel V., Hernquist L.,
  2013, Monthly Notices of the Royal Astronomical Society, 435, 1426

\bibitem[{Genel {et~al}\mbox{.}(2014)Genel, Vogelsberger, Springel, Sijacki,
  Nelson, Snyder, Rodriguez-Gomez, Torrey, \& Hernquist}]{Genel2014a}
Genel S. {et~al.}, 2014, Monthly Notices of the Royal Astronomical Society,
  445, 175

\bibitem[{Guo \& White(2008)}]{Guo2008}
Guo Q., White S. D.~M., 2008, Monthly Notices of the Royal Astronomical
  Society, 384, 2

\bibitem[{Hayward {et~al}\mbox{.}(2014)Hayward, Torrey, Springel, Hernquist, \&
  Vogelsberger}]{Hayward2014}
Hayward C.~C., Torrey P., Springel V., Hernquist L., Vogelsberger M., 2014,
  Monthly Notices of the Royal Astronomical Society, 442, 1992

\bibitem[{Hinshaw {et~al}\mbox{.}(2013)Hinshaw, Larson, Komatsu, Spergel,
  Bennett, Dunkley, Nolta, Halpern, Hill, Odegard, Page, Smith, Weiland, Gold,
  Jarosik, Kogut, Limon, Meyer, Tucker, Wollack, \& Wright}]{Hinshaw2013}
Hinshaw G. {et~al.}, 2013, The Astrophysical Journal Supplement Series, 208, 19

\bibitem[{Hopkins {et~al}\mbox{.}(2010{\natexlab{a}})Hopkins, Bundy, Croton,
  Hernquist, Keres, Khochfar, Stewart, Wetzel, \& Younger}]{Hopkins2010}
Hopkins P.~F. {et~al.}, 2010{\natexlab{a}}, The Astrophysical Journal, 715, 202

\bibitem[{Hopkins {et~al}\mbox{.}(2010{\natexlab{b}})Hopkins, Croton, Bundy,
  Khochfar, van~den Bosch, Somerville, Wetzel, Keres, Hernquist, Stewart,
  Younger, Genel, \& Ma}]{Hopkins2010a}
Hopkins P.~F. {et~al.}, 2010{\natexlab{b}}, The Astrophysical Journal, 724, 915

\bibitem[{Kartaltepe {et~al}\mbox{.}(2007)Kartaltepe, Sanders, Scoville,
  Calzetti, Capak, Koekemoer, Mobasher, Murayama, Salvato, Sasaki, \&
  Taniguchi}]{Kartaltepe2007}
Kartaltepe J.~S. {et~al.}, 2007, The Astrophysical Journal Supplement Series,
  172, 320

\bibitem[{Kaviraj {et~al}\mbox{.}(2014)Kaviraj, Devriendt, Dubois, Slyz,
  Welker, Pichon, Peirani, \& Borgne}]{Kaviraj2014}
Kaviraj S., Devriendt J., Dubois Y., Slyz A., Welker C., Pichon C., Peirani S.,
  Borgne D.~L., 2014, preprint (arXiv:1411.2595)

\bibitem[{Lacey \& Cole(1993)}]{Lacey1993}
Lacey C., Cole S., 1993, Monthly Notices of the Royal Astronomical Society,
  262, 627

\bibitem[{Lackner {et~al}\mbox{.}(2014)Lackner, Silverman, Salvato, Kampczyk,
  Kartaltepe, Sanders, Capak, Civano, Halliday, Ilbert, Jahnke, Koekemoer, Lee,
  {Le F\`{e}vre}, Liu, Scoville, Sheth, \& Toft}]{Lackner2014}
Lackner C.~N. {et~al.}, 2014, The Astronomical Journal, 148, 137

\bibitem[{Lee {et~al}\mbox{.}(2014)Lee, Yi, Elahi, Thomas, Pearce, Behroozi,
  Han, Helly, Jung, Knebe, Mao, Onions, Rodriguez-Gomez, Schneider, Srisawat,
  \& Tweed}]{Lee2014}
Lee J. {et~al.}, 2014, Monthly Notices of the Royal Astronomical Society, 445,
  4197

\bibitem[{Lemson \& Springel(2006)}]{Lemson2006}
Lemson G., Springel V., 2006, Astronomical Data Analysis Software and Systems
  XV ASP Conference Series, 351, 212

\bibitem[{Lin {et~al}\mbox{.}(2008)Lin, Patton, Koo, Casteels, Conselice,
  Faber, Lotz, Willmer, Hsieh, Chiueh, Newman, Novak, Weiner, \&
  Cooper}]{Lin2008}
Lin L. {et~al.}, 2008, The Astrophysical Journal, 681, 232

\bibitem[{L\'{o}pez-Sanjuan {et~al}\mbox{.}(2012)L\'{o}pez-Sanjuan, {Le
  F\`{e}vre}, Ilbert, Tasca, Bridge, Cucciati, Kampczyk, Pozzetti, Xu, Carollo,
  Contini, Kneib, Lilly, Mainieri, Renzini, Sanders, Scodeggio, Scoville,
  Taniguchi, Zamorani, Aussel, Bardelli, Bolzonella, Bongiorno, Capak, Caputi,
  de~la Torre, de~Ravel, Franzetti, Garilli, Iovino, Knobel, Kova\v{c},
  Lamareille, {Le Borgne}, {Le Brun}, {Le Floc’h}, Maier, McCracken, Mignoli,
  Pell\'{o}, Peng, P\'{e}rez-Montero, Presotto, Ricciardelli, Salvato,
  Silverman, Tanaka, Tresse, Vergani, Zucca, Barnes, Bordoloi, Cappi, Cimatti,
  Coppa, Koekemoer, Liu, Moresco, Nair, Oesch, Schawinski, \&
  Welikala}]{Lopez-Sanjuan2012}
L\'{o}pez-Sanjuan C. {et~al.}, 2012, Astronomy \& Astrophysics, 548, A7

\bibitem[{Lotz {et~al}\mbox{.}(2011)Lotz, Jonsson, Cox, Croton, Primack,
  Somerville, \& Stewart}]{Lotz2011}
Lotz J.~M., Jonsson P., Cox T.~J., Croton D., Primack J.~R., Somerville R.~S.,
  Stewart K., 2011, The Astrophysical Journal, 742, 103

\bibitem[{Maller {et~al}\mbox{.}(2006)Maller, Katz, Kere\v{s}, Dave, \&
  Weinberg}]{Maller2006}
Maller A.~H., Katz N., Kere\v{s} D., Dave R., Weinberg D.~H., 2006, The
  Astrophysical Journal, 647, 763

\bibitem[{Man {et~al}\mbox{.}(2012)Man, Toft, Zirm, Wuyts, \& van~der
  Wel}]{Man2012}
Man A. W.~S., Toft S., Zirm A.~W., Wuyts S., van~der Wel A., 2012, The
  Astrophysical Journal, 744, 85

\bibitem[{Man, Zirm \& Toft(2014)Man, Zirm, \& Toft}]{Man2014}
Man A. W.~S., Zirm A.~W., Toft S., 2014, preprint (arXiv:1410.3479)

\bibitem[{Marinacci, Pakmor \& Springel(2013)Marinacci, Pakmor, \&
  Springel}]{Marinacci2013}
Marinacci F., Pakmor R., Springel V., 2013, Monthly Notices of the Royal
  Astronomical Society, 437, 1750

\bibitem[{McMillan(2011)}]{McMillan2011}
McMillan P.~J., 2011, Monthly Notices of the Royal Astronomical Society, 414,
  2446

\bibitem[{Moreno {et~al}\mbox{.}(2013)Moreno, Bluck, Ellison, Patton, Torrey,
  \& Moster}]{Moreno2013}
Moreno J., Bluck a. F.~L., Ellison S.~L., Patton D.~R., Torrey P., Moster
  B.~P., 2013, Monthly Notices of the Royal Astronomical Society, 436, 1765

\bibitem[{Muldrew, Pearce \& Power(2011)Muldrew, Pearce, \&
  Power}]{Muldrew2011}
Muldrew S.~I., Pearce F.~R., Power C., 2011, Monthly Notices of the Royal
  Astronomical Society, 410, 2617

\bibitem[{Neistein \& Dekel(2008{\natexlab{a}})}]{Neistein2008a}
Neistein E., Dekel A., 2008{\natexlab{a}}, Monthly Notices of the Royal
  Astronomical Society, 383, 615

\bibitem[{Neistein \& Dekel(2008{\natexlab{b}})}]{Neistein2008}
Neistein E., Dekel A., 2008{\natexlab{b}}, Monthly Notices of the Royal
  Astronomical Society, 388, 1792

\bibitem[{Nelson {et~al}\mbox{.}(2013)Nelson, Vogelsberger, Genel, Sijacki,
  Keres, Springel, \& Hernquist}]{Nelson2013b}
Nelson D., Vogelsberger M., Genel S., Sijacki D., Keres D., Springel V.,
  Hernquist L., 2013, Monthly Notices of the Royal Astronomical Society, 429,
  3353

\bibitem[{Newman {et~al}\mbox{.}(2012)Newman, Ellis, Bundy, \&
  Treu}]{Newman2012}
Newman A.~B., Ellis R.~S., Bundy K., Treu T., 2012, The Astrophysical Journal,
  746, 162

\bibitem[{Press \& Schechter(1974)}]{Press1974}
Press W.~H., Schechter P., 1974, {Formation of Galaxies and Clusters of
  Galaxies by Self-Similar Gravitational Condensation}

\bibitem[{Sales {et~al}\mbox{.}(2007)Sales, Navarro, Abadi, \&
  Steinmetz}]{Sales2007}
Sales L.~V., Navarro J.~F., Abadi M.~G., Steinmetz M., 2007, Monthly Notices of
  the Royal Astronomical Society, 379, 1464

\bibitem[{Sales {et~al}\mbox{.}(2015)Sales, Vogelsberger, Genel, Torrey,
  Nelson, Rodriguez-Gomez, Wang, Pillepich, Sijacki, Springel, \&
  Hernquist}]{Sales2014a}
Sales L.~V. {et~al.}, 2015, Monthly Notices of the Royal Astronomical Society:
  Letters, 447, L6

\bibitem[{Somerville {et~al}\mbox{.}(2008)Somerville, Hopkins, Cox, Robertson,
  \& Hernquist}]{Somerville2008}
Somerville R.~S., Hopkins P.~F., Cox T.~J., Robertson B.~E., Hernquist L.,
  2008, Monthly Notices of the Royal Astronomical Society, 391, 481

\bibitem[{Springel(2010)}]{Springel2010}
Springel V., 2010, Monthly Notices of the Royal Astronomical Society, 401, 791

\bibitem[{Springel \& Hernquist(2003)}]{Springel2003}
Springel V., Hernquist L., 2003, Monthly Notices of the Royal Astronomical
  Society, 339, 289

\bibitem[{Springel {et~al}\mbox{.}(2005)Springel, White, Jenkins, Frenk,
  Yoshida, Gao, Navarro, Thacker, Croton, Helly, Peacock, Cole, Thomas,
  Couchman, Evrard, Colberg, \& Pearce}]{Springel2005b}
Springel V. {et~al.}, 2005, Nature, 435, 629

\bibitem[{Springel {et~al}\mbox{.}(2001)Springel, White, Tormen, \&
  Kauffmann}]{Springel2001}
Springel V., White S. D.~M., Tormen G., Kauffmann G., 2001, Monthly Notices of
  the Royal Astronomical Society, 328, 726

\bibitem[{Srisawat {et~al}\mbox{.}(2013)Srisawat, Knebe, Pearce, Schneider,
  Thomas, Behroozi, Dolag, Elahi, Han, Helly, Jing, Jung, Lee, Mao, Onions,
  Rodriguez-Gomez, Tweed, \& Yi}]{Srisawat2013a}
Srisawat C. {et~al.}, 2013, Monthly Notices of the Royal Astronomical Society,
  436, 150

\bibitem[{Stewart {et~al}\mbox{.}(2009)Stewart, Bullock, Barton, \&
  Wechsler}]{Stewart2009}
Stewart K.~R., Bullock J.~S., Barton E.~J., Wechsler R.~H., 2009, The
  Astrophysical Journal, 702, 1005

\bibitem[{Torrey {et~al}\mbox{.}(2014{\natexlab{a}})Torrey, Snyder,
  Vogelsberger, Hayward, Genel, Sijacki, Springel, Hernquist, Nelson, Kriek,
  Pillepich, Sales, \& McBride}]{Torrey2014a}
Torrey P. {et~al.}, 2014{\natexlab{a}}, preprint (arXiv:1411.3717)

\bibitem[{Torrey {et~al}\mbox{.}(2014{\natexlab{b}})Torrey, Vogelsberger,
  Genel, Sijacki, Springel, \& Hernquist}]{Torrey2014}
Torrey P., Vogelsberger M., Genel S., Sijacki D., Springel V., Hernquist L.,
  2014{\natexlab{b}}, Monthly Notices of the Royal Astronomical Society, 438,
  1985

\bibitem[{Vogelsberger {et~al}\mbox{.}(2013)Vogelsberger, Genel, Sijacki,
  Torrey, Springel, \& Hernquist}]{Vogelsberger2013}
Vogelsberger M., Genel S., Sijacki D., Torrey P., Springel V., Hernquist L.,
  2013, Monthly Notices of the Royal Astronomical Society, 436, 3031

\bibitem[{Vogelsberger {et~al}\mbox{.}(2014{\natexlab{a}})Vogelsberger, Genel,
  Springel, Torrey, Sijacki, Xu, Snyder, Bird, Nelson, \&
  Hernquist}]{Vogelsberger2014a}
Vogelsberger M. {et~al.}, 2014{\natexlab{a}}, Nature, 509, 177

\bibitem[{Vogelsberger {et~al}\mbox{.}(2014{\natexlab{b}})Vogelsberger, Genel,
  Springel, Torrey, Sijacki, Xu, Snyder, Nelson, \&
  Hernquist}]{Vogelsberger2014}
Vogelsberger M. {et~al.}, 2014{\natexlab{b}}, Monthly Notices of the Royal
  Astronomical Society, 444, 1518

\bibitem[{Vogelsberger {et~al}\mbox{.}(2012)Vogelsberger, Sijacki, Kere\v{s},
  Springel, \& Hernquist}]{Vogelsberger2012}
Vogelsberger M., Sijacki D., Kere\v{s} D., Springel V., Hernquist L., 2012,
  Monthly Notices of the Royal Astronomical Society, 425, 3024

\bibitem[{Vogelsberger {et~al}\mbox{.}(2014{\natexlab{c}})Vogelsberger, Zavala,
  Simpson, \& Jenkins}]{Vogelsberger2014c}
Vogelsberger M., Zavala J., Simpson C., Jenkins a., 2014{\natexlab{c}}, Monthly
  Notices of the Royal Astronomical Society, 444, 3684

\bibitem[{Wetzel, Cohn \& White(2009)Wetzel, Cohn, \& White}]{Wetzel2009a}
Wetzel A.~R., Cohn J.~D., White M., 2009, Monthly Notices of the Royal
  Astronomical Society, 395, 1376

\bibitem[{Williams, Quadri \& Franx(2011)Williams, Quadri, \&
  Franx}]{Williams2011}
Williams R.~J., Quadri R.~F., Franx M., 2011, The Astrophysical Journal, 738, 7

\bibitem[{Xu {et~al}\mbox{.}(2012)Xu, Zhao, Scoville, Capak, Drory, \&
  Gao}]{Xu2012}
Xu C.~K., Zhao Y., Scoville N., Capak P., Drory N., Gao Y., 2012, The
  Astrophysical Journal, 747, 85

\end{thebibliography}

\end{document}